\documentclass{ar-1col}
\usepackage{url}
\usepackage{comment} 

\usepackage[caption=false]{subfig}

\usepackage{natbib}
\usepackage{amsmath}

\usepackage{graphicx}

\usepackage[margin=1in]{geometry}
\jname{Annu. Rev. Astron. Astrophys.}
\jvol{63}
\jyear{2025}
\doi{10.1146/annurev-astro-121024-051405}

\begin{document}

\markboth{Strigari}{Timing Mass of the Local Group}

\title{Timing Mass of the Local Group}

\author{Louis E. Strigari,$^1$ 
\affil{$^1$Department of Physics and Astronomy, Mitchell Institute
  for Fundamental Physics and Astronomy, Texas A\&M University,
  College Station, Texas, USA; email: strigari@tamu.edu}
}

\begin{abstract}
The classic model of the Local Group (LG) is that of two dominant constituents, the Milky Way and M31, first separating and then detaching from the Hubble flow, leading to a nearly radial approaching orbit. This simple model has been confronted by new measurements of the 3D M31 kinematics, by cosmological simulations, and by theoretical understanding of the impact of massive substructures such as the Large Magellanic Cloud. This article explores the consequences of new observations and theory on the determination of the mass and dynamics of the LG. 
The M31 tangential velocity measurement and contribution from the cosmological constant both increase the implied timing mass of the LG to be $\sim 5 \times 10^{12}$ M$_\odot$. Timing mass estimates for the LG tend to be larger than the sum of the Milky Way and M31 halo masses, and larger than independent LG mass estimators. Precision future kinematics have the potential to explore the origin of this difference, shed light on dark matter in the LG, the origin of its angular momentum, and possibly even local values of cosmological parameters. 
\end{abstract}

\maketitle

\tableofcontents

\section{Introduction}

\par The discovery of galaxies beyond the realm of our Milky Way (MW) established a new perspective on the scale of the Universe, and our place within it. Amongst all of the island universes beyond the MW, the Andromeda galaxy, or M31, stands out as especially significant. 
M31 is the nearest galaxy with properties similar to the MW. On the Hubble sequence, M31 is classified as an SA(s)b spiral galaxy, similar to the traditional classification of the MW as an Sb-Sbc galaxy. The MW and M31 are cosmic siblings, whose pasts and futures are intertwined. 

\par M31 has been studied across the electromagnetic spectrum.~\cite{Babcock:1939} was among the first to measure the M31 rotation curve, and from this estimate also came its mass distribution. Using radial velocity measurements in the optical emission line region and extending beyond the M31 stellar disk, Babcock determined that the total mass-to-light ratio increases in the outer regions. The increasing M31 mass-to-light ratio was corroborated by 21-cm observations~\citep{vandeHulst1957,Roberts1975}. These results provided  evidence for a flat M31 rotation curve, disfavoring a Keplerian fall-off as expected for gravitational dynamics dominated by the visible stars and gas. The measurement of a flat rotation curve for M31 was similar to that obtained from a larger sample of nearby spiral galaxies ~\citep{Bosma1978,Rubin1980}. Accounting for systematics, such as the effect of beam smearing, flat rotation curves were measured in the outermost regions of many spiral galaxies~\citep{Sofue:2000jx}. Clear evidence for an extended dark matter (DM) halo around M31 was thereby established.  

\par Similar to M31 and the nearby spiral galaxies, evidence pointed towards a flat rotation curve for the MW.~\cite{Kinman:1959} used the kinematics of globular clusters and satellite galaxies to measure the mass of the MW beyond the visible disk. This analysis was extended with improved observational samples and more detailed theoretical models~\citep{Little:1987,Zaritsky:1989,Kulessa:1992,Kochanek:1995xv}. As in the case of M31, a variety of modern measurements provide clear evidence for an extended DM halo around the MW~\citep{2003A&A...397..899S,SDSS:2008nmx,2012ApJ...759..131B,Deason:2012wm}. 

\par The MW and M31 are the most prominent of the nearby galaxies, but by number they represent a small fraction of the galaxies in the Local Group (LG). The LG is defined as the approximately spherical volume with radius $\sim 1$ Mpc centered on the center of mass, or barycenter, of the MW-M31 system. Dwarf galaxies, which range from nearly 4 to 20 magnitudes fainter than the MW and M31, dominate the census of galaxies in the LG~\citep{2012AJ....144....4M}. There are nearly 80 such dwarf galaxies known, of which the large majority are satellites orbiting either the MW or M31. 

\par Though they dominate the number of LG galaxies, and individually they are amongst the most DM-dominated galaxies in the Universe~\citep{1983ApJ...266L..11A,1998ARA&A..36..435M,Simon:2007dq,2007NuPhS.173...15G,Strigari:2008ib,Wolf:2009tu}, dwarf galaxies constitute a small fraction of the LG mass. The most luminous such galaxies include the Large Magellanic Cloud (LMC) and M33, which are associated with the MW and M31, respectively. This large population of dwarf galaxies orbiting the MW and M31 is in line with expectations from the theory of cold dark matter (CDM)~\citep{Frenk:2012ph,Sawala:2014baa}, though many questions still linger on the details of matching the population of galaxies to theory predictions~\citep{Boylan-Kolchin:2011lmk,2017ARA&A..55..343B}. 

\par The LG consists of the MW, M31, and the population of dwarf galaxies, but is this all there is to the LG? A question similar in spirit was posed over a half century ago by~\cite{1959ApJ...130..705K}, following up on initial measurements of the relative motion of the MW and M31. These authors made the simple assumption that the MW and M31 dominate the mass of the LG, that their two-body dynamics is described by a Keplerian orbit, and that these two galaxies are approaching one another on a nearly radial orbit. If in the distant past the two galaxies were receding due to the Hubble expansion, at the time of the Big Bang they could be modeled as being nearly spatially coincident. Given these assumptions,~\cite{1959ApJ...130..705K} found that there was at least ten times more unseen material within both of the galaxies. It is interesting to note that this unseen mass was not viewed in the context of DM, but rather these authors suggested that this material was in the form of intergalactic gas and dust that may be revealed in nearby emission~\citep{1999ApJ...522L..81M}. In retrospect, this estimate provided one of the first hints for DM in and around the MW and M31, and established a connection between the dynamics of local galaxies to the universe on larger scales. 
  
\par As it was derived from a simple model for the LG, this mass estimate has become known as the timing argument, and throughout this article it will be referred to as the timing mass of the LG. Despite the simplicity of the assumptions, the result has stood the test of time, perhaps because the simplified model of the orbit is appropriate. Indeed, ~\cite{1974CoASP...6....7G} motivates a purely radial MW-M31 orbit on first infall, because there are no tidal shocks observed in the LG that would indicate a previous close passage between the galaxies.~\cite{1982MNRAS.199...67E} first presented indirect evidence that the MW-M31 orbit is not purely radial, though a tangential component to the velocity could not be conclusively established. 

\par Over the subsequent years, the structure and kinematics of the MW, M31, and LG have come into sharper focus~\citep{1998ARA&A..36..435M}. There are now several dozen known dwarf galaxies in the LG, over half of which have been discovered in the last two decades~\citep{2010AdAst2010E..21W,2019ARA&A..57..375S}. The internal structure of the MW and M31 have been revealed by large-scale photometric surveys~\citep{2008ApJ...673..864J}, and the Gaia observatory~\citep{2021A&A...649A...1G}. Perhaps the most important recent development regarding LG kinematics is the measurement of the M31 proper motion. In particular, {\it Hubble Space Telescope} (HST)~\citep{2012ApJ...753....7S} and {\it Gaia}~\citep{2021MNRAS.507.2592S} data provide evidence for a non-zero MW-M31 relative tangential velocity, which has important implications for not only the timing mass, but also for the formation and future evolution of the LG~\citep{2024arXiv240800064S}. 

\par In parallel with the kinematic measurements, simulations of LG-like systems have developed as a powerful theoretical tool. Simulations provide predictions for how much of the LG mass is contained within the two main galaxies, within the dwarf galaxies, and also within dark substructure that does not host visible galaxies. They also provide a sense of whether the census of LG dwarf galaxies is complete, and how these dwarf galaxies are distributed in the LG. From the kinematic perspective, they can determine how likely it is that a MW-M31 like system has acquired a significant tangential velocity.  

\par With the unprecedented observations and theoretical tools now at our disposal, we are in position to answer two fundamental questions on the LG: 1) How is the dark and luminous mass distributed, and 2) What is the nature of the MW-M31 orbit? The first question may be addressed by comparing the timing mass, as well as independent LG mass estimates, to the sum of the masses of the known constituents of the LG. Though this is straightforward in principle, this comparison is complicated by systematics in the mass estimator algorithms. Accounting for such systematics as carefully as possible, timing mass estimates typically find a larger LG mass than the directly measured MW and M31 mass sum. 

\par The second question on the nature of the MW-M31 orbit may be answered via more precise kinematic measurements, and sheds light on the formation history of the LG. This question is also informed by LG simulations which are able to model the assembly history of the LG. Given the recent measurements, is the paradigm shifting from a pure radial Keplerian orbit, to a more complicated orbit in which tangential velocity and dark energy must be taken into account? Additionally, what is the impact of DM substructure and dwarf galaxies on the LG kinematics? 

\par The primary aim of this review is to address the two questions above. In the process of exploring the answers to these questions, the unique connection that the LG provides to cosmology and the properties of the Universe is highlighted. Indeed the boundary of the LG is the nearest point at which the cosmological expansion becomes relevant. The physical properties of the LG should be connected to the environment outside of the LG, and can even provide a means to measure fundamental cosmological parameters on the smallest scales possible. 

\section{Local Group Constituents}
\par The aim of this section is to define convenient coordinate systems used to describe the LG and its constituents. The basic properties of the main LG constituents, the MW and M31, are then reviewed, followed by a discussion of the observed dwarf galaxies and the predictions for DM substructure within the LG. 

\subsection{Coordinate systems}
\par In the Galactic coordinate system, the direction to the center of the MW disk defines latitude and longitude $(\ell, b) = (0,0)$ degrees. The Galactocentric cartesian coordinate system is defined such that the center of the MW disk is located at the origin, $(x,y,z) = (0,0,0)$. The Sun is located along the $x$-axis at a distance of $x = -R_\odot$, where $R_\odot$ is the distance from the Sun to the Galactic center. The $y$-axis points in the direction of the rotation of the Galactic disk, and the $z$-axis points towards the North Galactic pole.  
Defining a unit vector along a given line-of-sight as $\hat \rho$, and similarly unit vectors along longitude and latitude as $(\hat \ell, \hat b)$, this implies that the unit vector from the Sun in the direction $(\ell,b)$ is  
\begin{equation} 
    \hat \rho = (\cos \ell \cos b, \sin \ell \cos b, \sin b).  
\label{eq:lunit} 
\end{equation}

\par The LG barycenteric coordinate system has its origin at the MW-M31 center-of-mass, assuming that these two galaxies dominate the mass of the LG. The exact location of this origin is unknown and depends on the ratio of the MW and M31 mass. The measured dipole anisotropy of the Cosmic Microwave Background (CMB) is likely a result of the motion of the LG barycenter relative to the isotropic CMB frame. The best fit value for the magnitude of this velocity is~\citep{2020A&A...641A...1P}
\begin{equation} 
v_{LG \rightarrow CMB} = 620 \pm 15 \, {\rm km} \, \rm{s}^{-1}
\label{eq:VLGCMB}
\end{equation} 
in the direction $\ell = 271.9 \pm 2.0$ and $b = 29.6 \pm 1.4$ degrees.

\par The Galactocentric cartesian coordinate system may be converted into a {\it supergalactic} coordinate system, which has components labeled as $(SGX,SGY,SGZ)$. On scales larger than the LG but still within the local volume, defined roughly as a sphere of radius approximately $10$ Mpc, matter is distributed along a flat, sheet-like structure. In the supergalactic coordinate system, the plane of the local supercluster is approximately aligned with the $(SGX, SGY)$ plane. The MW disk is nearly perpendicular to the supergalactic plane, with the supergalactic north pole located at Galactic coordinates of $(\ell,b) = (47.37,6.32)$ degrees. 

\subsection{Milky Way} 
\par The MW is the best-studied of all galaxies~\citep{2016ARA&A..54..529B}. The three primary baryonic components of the MW are the bar/bulge, the disk, and the stellar halo. By mass, the dominant component of the disk is the thin disk. The scaleheight of the thin disk perpendicular to the Galactic plane is approximately $300$ pc, and the total mass of the thin disk is approximately $5 \times 10^{10}$ M$_\odot$. The second component is the thick disk, which has a scaleheight of approximately $1$ kpc and is composed of older stars that are more metal-poor than the thin disk. By mass, about 90\% of the visible material is contained in stars, whereas the remaining 10\% is in the form of gas and dust. The total absolute visual magnitude of the MW is estimated to be $M_V = -20.9$, corresponding to a total luminosity of approximately a few times $10^{10}$ L$_\odot$. 

\par The Sun is located at a distance of $R_\odot = 8.2$ kpc from the Galactic center~\citep{2009ApJ...707L.114G}. At the position of the Sun, the circular velocity is 239 km/s, which in Galactocentric coordinates corresponds to a Local Standard of Rest (LSR) velocity of~\citep{2011MNRAS.414.2446M} 
\begin{equation}
{\vec v}_{LSR \rightarrow MW} = (0,239,0) \, {\rm km/s}. 
\label{eq:vlsrmw}
\end{equation} 
The mean velocity of the Sun with respect to the LSR is  
\begin{equation} 
{\vec v}_{\odot \rightarrow LSR} = (11.1,12.24,7.25) \, {\rm km/s}.
\label{eq:vlsr}
\end{equation} 

\par Beyond the visible disk extends the MW DM halo, which is the dominant component of mass for the Galaxy. The outer edge, corresponding to the approximate virial radius, of the MW DM halo is estimated to be $300$ kpc~\citep{2020MNRAS.496.3929D}. Below independent measurements for the mass of the MW DM halo are reviewed.  

\subsection{M31}
\par M31 is unique in that it is the only galaxy similar to the MW that may be studied via its resolved stellar populations. Also since it is viewed from the outside, it provides a unique perspective on a galaxy similar to the MW. The absolute B-band magnitude for M31 is $-20.81$~\citep{1988A&A...198...61W}, and the total stellar mass is $\sim (10-15) \times 10^{10}$ $M_\odot$~\citep{2012A&A...546A...4T}. The M31 disk subtends an angle of $1.8^\circ$, the inclination of the disk is $77^\circ$ and the position angle is $37.5^\circ$. In the outermost regions, the circular velocity peaks at $\sim 275$ km/s~\citep{2009ApJ...705.1395C}. 

\par Though it is similar in many respects to the MW, recent observations highlight interesting differences. For example, there is evidence that M31 has undergone several significant mergers within the past 3-4 Gyr~\citep{2018ApJ...868...55M}. In contrast, the last significant merger for the MW happened nearly 8 Gyr ago~\citep{2018Natur.563...85H,2020MNRAS.494.3880B}. 

\par There is evidence for an extended DM halo from the motion of the outer halo stars and satellite galaxies. The outer edge of the M31 halo is at $\sim 300$ kpc, nearly overlapping with the edge of the MW halo. DM halo mass estimates for M31 are reviewed below. 

\subsection{Dwarf galaxies and dark matter substructure}

\par In addition to the MW and M31, the LG contains a population of dwarf galaxies. Very broadly these dwarf galaxies can be classified as satellites of the MW, satellites of M31, or ``field" galaxies that are not bound to the MW or to M31. The most luminous galaxy is the Large Magellanic Cloud (LMC), which is classified as a dwarf irregular galaxy that is two magnitudes fainter that the MW. The LMC and Small Magellanic Cloud (SMC) are satellite galaxies of the MW. Modeling of its orbit indicates that the LMC is infalling into the MW for the first time~\citep{2007ApJ...668..949B,2009ApJ...700..924K,2011ApJ...742..110N,2023Galax..11...59V}, and it is expected that the many satellite galaxies are associated with the LMC~\citep{Sales:2011id}. There are approximately 60 galaxies classified as part of the MW subgroup,  with the majority of these likely bound to the MW as satellite galaxies~\citep{2012AJ....144....4M}. Similarly, there are 34 galaxies classified as part of the M31 subgroup, with the majority of these likely bound to M31 as satellite galaxies. Properties of MW satellites have been reviewed in depth~\citep{2019ARA&A..57..375S}. Here satellites not bound to either the MW or M31 are referred to as outer Local Group members (OLGMs). 

\par
\begin{figure}[h]
\centering 
\includegraphics[width=4.0in]{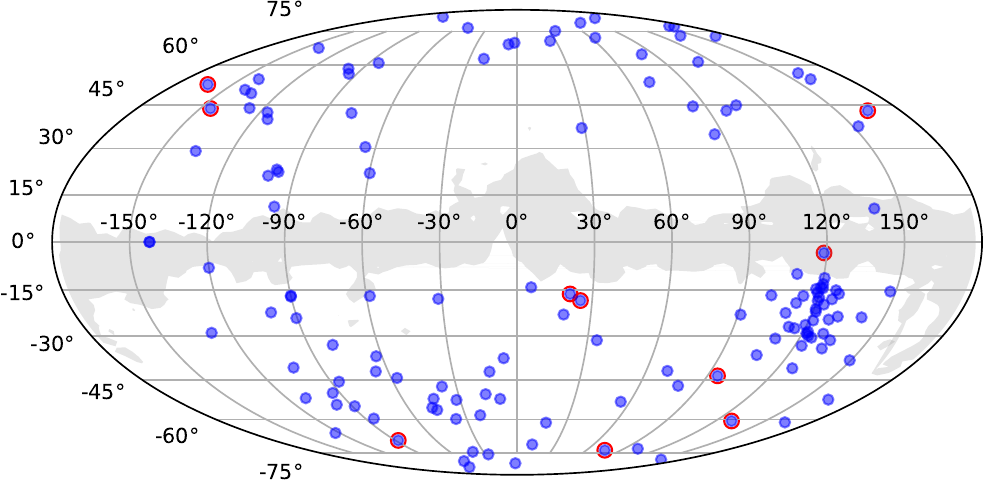} 
\caption{All-sky projection of Local Group galaxies ({\it blue points}) in galactic coordinates. Red circles indicate the Outer Local Group Members that are not satellites of the MW or M31. The cluster of points at the lower right are satellites of M31. The grey shaded regions indicate extinction with $E(B-V) > 0.3$ along the Galactic plane as taken from the 2MASS Redshift Survey~\citep{2012ApJS..199...26H}.
}\label{fig:LG}
\end{figure}

 Figure~\ref{fig:LG} shows the projected sky positions of LG members. MW satellites are distributed nearly uniformly across the sky. There are few satellites that coincide with the Galactic plane, $b < 15$ degrees, since faint satellites are obscured from detection at these latitudes. In Figure~\ref{fig:LG} a cluster of satellites is evident surrounding M31. OLGMs are also uniform, though most discovered to date are at negative galactic latitudes. Though there has been significant progress in the detection of faint satellite galaxies~\citep{DES:2015zwj,DES:2018gui}, selection effects persist. Surveys such as that by the upcoming Vera C. Rubin observatory are expected to improve the census of LG galaxies.  

\par Simulations predict the population of LG satellites galaxies~\citep{2024MNRAS.532.2490S}. In addition to luminous galaxies, simulations predict a substantial population of completely dark subhalos that do not host luminous galaxies~\citep{1999ApJ...522...82K,1999ApJ...524L..19M}. These systems may constitute a significant contribution to the mass of the LG. The mass of these subhalos could go down to Earth masses or even lower. 

\par As it is the most massive dwarf galaxy in the LG, the LMC is modeled to have a significant impact of the kinematics of the MW. As the LMC infalls into the MW halo, it creates a dipole in the nearby halo stars relative to the more distant outer halo stars, the latter of which have longer dynamical response times. As a result, the LMC impact induces a relative motion between the MW disk and the MW DM halo, resulting in a displacement in the velocities between the disk and the DM halo~\citep{2015ApJ...802..128G,2019ApJ...884...51G,2021Natur.592..534C,2021MNRAS.506.2677E}. The results from~\citet{2021NatAs...5..251P} show that the magnitude of this ``travel" velocity is $32$ km/s, in the direction of $(\ell, b) = (56,-34)$ degrees. 
 
\section{M31 spatial motion}
\label{sec:M31PM}
\par In measuring the displacement of the M31 nucleus relative to three nearby stars,~\cite{1917AJ.....30..175B} attempted the first measurement of the M31 proper motion. No discernible motion of M31 was measured during the 18 year observational period, and given the sensitivity this corresponded to a limit on the proper motion on the order of $\sim 1$ mas/yr. In terms of tangential velocities, using the relation $v_{\rm t} = 4.74 [ \mu / ({\rm mas/yr})] [D / ({\rm kpc)}]$ km/s, where $D = 770$ kpc is the distance to M31, this implied a bound on the tangential velocity of $v_{\rm t} < 3000$ km/s, much larger than the line-of-sight velocity measured at the time~\citep{1913LowOB...2...56S}. 

\par Over a century later, the M31 proper motion has now been established through substantial improvement in the astrometric precision and the resolution of member stars. This section reviews the astrometric measurements, highlighting in particular direct astrometric measurements from HST and {\it Gaia}. Complementing the direct astrometric measurements, independent measurements of the M31 tangential velocity come from the kinematics of the OLGMs and M31 satellites. These require assumptions on the dynamical state of M31 and the LG, and provide a cross-check on the direct proper motions from HST and {\it Gaia}. 

\par Determination of the M31 intrinsic proper motion or tangential velocity requires subtracting the velocity component that is due to the motion of the observer. This is known to result from the motion of the Sun with respect to the Galactic rest frame. More recently, it has been understood that there is an additional component resulting from the reflex motion due to the impact of the LMC on the MW disk. This section begins by defining the measured velocities, establishing a methodology for subtraction of observer-related motion, and then moves on to discuss the different M31 proper motion and tangential velocity measurements. 

\subsection{Velocity definitions}

\par Let's begin by defining the full three-dimensional velocity of M31 with respect to the Sun as ${\vec v}_{M31 \rightarrow \odot}$. This velocity includes the two components in the plane of the sky, and the line-of-sight velocity component. The line-of-sight velocity of M31 has long been well known from Doppler shifts, with modern measurements finding a value of $-301 \pm 1$ km/s~\citep{vanderMarel:2007yw}. The minus sign indicates that M31 is moving in the direction of the Sun. 

\par Though it represents just one velocity component, a simple constraint on the properties of the MW-M31 orbit may be motivated from the precise measurement of the line-of-sight heliocentric velocity. Taking the MW rotational velocity and the velocity of the Sun relative to the LSR from Equations~\ref{eq:vlsrmw} and~\ref{eq:vlsr}, respectively, the component of the Solar velocity in the direction of M31 is $191$ km/s. Therefore the remaining $-110$ km/s provides an estimate for the intrinsic radial component of the M31 velocity relative to the MW. This estimate does not include the travel velocity correction resulting from the LMC, which as shown below tends to increase the radial velocity. 

\par To obtain the expectation for the M31 proper motion, the Solar motion may similarly be projected onto the plane of the sky at the position of M31. Define the proper motion along the right ascension direction as $\mu_\alpha$, with $\mu_{\alpha \star} = \mu_\alpha \cos \delta$, and along the declination direction as $\mu_\delta$. Again using the Solar rotational and LSR velocities from Equations~\ref{eq:vlsrmw} and~\ref{eq:vlsr}, the projection corresponds to proper motions of $(\mu_{\alpha \star}, \mu_\delta) = (-0.04,0.02)$ mas/yr. Any additional M31 heliocentric proper motion is due to the intrinsic motion of M31. 

\begin{table}[h]
\tabcolsep7.5pt
\caption{M31 tangential velocity measurements via the different methods shown in column 1, from the references shown in column 2. For the top three rows, columns 3 and 4 show the measured proper motions, from which the tangential velocities are derived. Columns 5 and 6 show corresponding relative MW-M31 radial and tangential velocities. For the top 3 rows, the relative velocities are derived assuming a heliocentric radial velocity of $-301 \pm 1$ km/s. In row 4, the method uses outer Local Group members and assumes momentum conservation between M31 and the MW, and the M31 satellites method in row 5 assumes that the M31 satellites are dynamically well-mixed in the M31 halo. For the bottom two rows, the radial and tangential velocities are directly derived from the given analysis method.
Abbreviations: HST, {\it Hubble Space Telescope}; LG, Local Group; MW, Milky Way. } 
\label{tab:M31PMs}
\begin{center}
\begin{tabular}{@{}c|c|c|c|c|c@{}}
\hline
\hline
Method & Reference & $\mu_{\alpha \star}$ [$\mu$as/yr] & $\mu_\delta$ [$\mu$as/yr] & $v_{\rm r}$ [km/s] & $v_{\rm t}$ [km/s] \\
\hline
\hline
HST & ~\citet{vanderMarel:2012xp} & $34 \pm 10$  & $-32 \pm 12$ & $-109 \pm 1 $ & $55 \pm 25 $ \\
Gaia DR2 & ~\citet{2019ApJ...872...24V} & $65 \pm 18$ & $-57 \pm 15$ & $-109 \pm 1 $ & $170 \pm 51 $ \\
Gaia EDR3 & ~\citet{2021MNRAS.507.2592S} & $49 \pm 11$ & $-38 \pm 8$ & $-109 \pm 1 $ & $78 \pm 28 $ \\
\hline 
Outer LG members & ~\citet{2014MNRAS.443.1688D} & -- & --& $-155 \pm 40$ & $114 \pm 39$  \\
M31 satellites & ~\citet{2016MNRAS.456.4432S} & -- & -- & $-88 \pm 14$ & $163 \pm 54$ \\
\hline
\end{tabular}
\end{center}
\end{table}

\par The relative M31-MW velocity may be written as  
\begin{equation} 
{\vec v}_{M31 \rightarrow MW} = {\vec v}_{M31 \rightarrow \odot} + {\vec v}_{\odot \rightarrow MW}. 
\label{eq:VM31MW}
\end{equation}
The second term on the right-hand side of Equation~\ref{eq:VM31MW} is the sum of the velocity of the Sun with respect to the LSR and the velocity of the LSR with respect to the MW, 
\begin{equation}
{\vec v}_{\odot \rightarrow MW} =  {\vec v}_{\odot \rightarrow LSR} + {\vec v}_{LSR \rightarrow MW}. 
\label{eq:VodotMW}
\end{equation}
with ${\vec v}_{LSR \rightarrow MW}$ and ${\vec v}_{\odot \rightarrow LSR}$ given in Equations~\ref{eq:vlsrmw} and~\ref{eq:vlsr}, respectively. Equation~\ref{eq:VodotMW} may be modified to include a contribution from the relative motion between the MW disk and the MW DM halo resulting from the LMC, as described above. Including this effect, Equation~\ref{eq:VodotMW} can be interpreted as the velocity of the Sun with respect to the MW disk, ${\vec v}_{\odot \rightarrow MW, disk}$, and a term which represents this travel velocity may be added, ${\vec v}_{MW, disk \rightarrow MW, halo}$, to obtain the full expression for ${\vec v}_{\odot \rightarrow MW}$. 

\par The relative radial velocity, $v_{\rm r}$, and relative tangential velocity, $v_{\rm t}$, components of ${\vec v}_{M31 \rightarrow MW}$ may be related to their Galactic cartesian components. The relationship between these two coordinate systems is derived using the MW-M31 separation in Galatocentric coordinates  
\begin{equation}
{\vec r}= 
( x, y, z ) = (-379.2, 612.7, -283.1) \, {\rm kpc}. 
~\label{eq:MWM31separation}
\end{equation}
Now, we define the relative velocity components in Galatocentric cartesian coordinates as $\vec v = (v_x,v_y,v_z)$, and the corresponding velocity in spherical coordinates $\vec v = (v_{\rm r},v_\phi,v_\theta)$, where $\phi$ may be measured from the positive $x-$axis, and $\theta$ is the polar angle measured from the $z-$axis. The spherical velocity components are then derived in a standard manner via projection of unit vectors in the coordinate systems. The relative tangential velocity is then $v_{\rm t}^2 = v_\theta^2 + v_\phi^2$. 

\subsection{{\it Hubble Space Telescope} and {\it Gaia}}
\par Using the HST ACS/WFC (Advanced Camera for Surveys/Wide Field Channel) and WFC3/UVIS (Wide Field Camera 3/Ultraviolet and Visible Light Channel) instruments,~\cite{2012ApJ...753....7S} established the first modern M31 proper motion measurement. They examined three fields in the vicinity of the M31 disk, measuring the astrometry of thousands of stars on a time baseline of 5-7 years. These astrometry measurements obtained via HST are differential, in the sense that the positional measurements are obtained in a small area of sky relative to comparison field objects. For the M31 measurement, the comparison objects were galaxies, which significantly outnumber other possible comparison objects, such as quasars. For the three fields considered,~\citet{vanderMarel:2012xp} correct the results of~\cite{2012ApJ...753....7S} for internal motions to obtain a weighted average proper motion of $(\mu_{\alpha \star}, \mu_\delta) = (34 \pm 10,-32 \pm 12)$ $\mu$as/yr. 
  
\par Launched in 2013, the {\it Gaia} satellite measures precision astrometry for over one billion sources~\citep{Gaia:2018ydn}. {\it Gaia} is a full-sky survey, measuring parallaxes and proper motions from large-scale, global absolute observations. Amongst its many contributions, {\it Gaia} has provided important new insight into LG galaxies, including measurement of their proper motions, from which orbital trajectories and infall times for satellites into the MW have been determined~\citep{Gaia:2018kkg,2018A&A...619A.103F,2018ApJ...863...89S}. 

\par
\begin{figure}
\begin{minipage}{.45\textwidth}
    {\includegraphics[width=\textwidth]{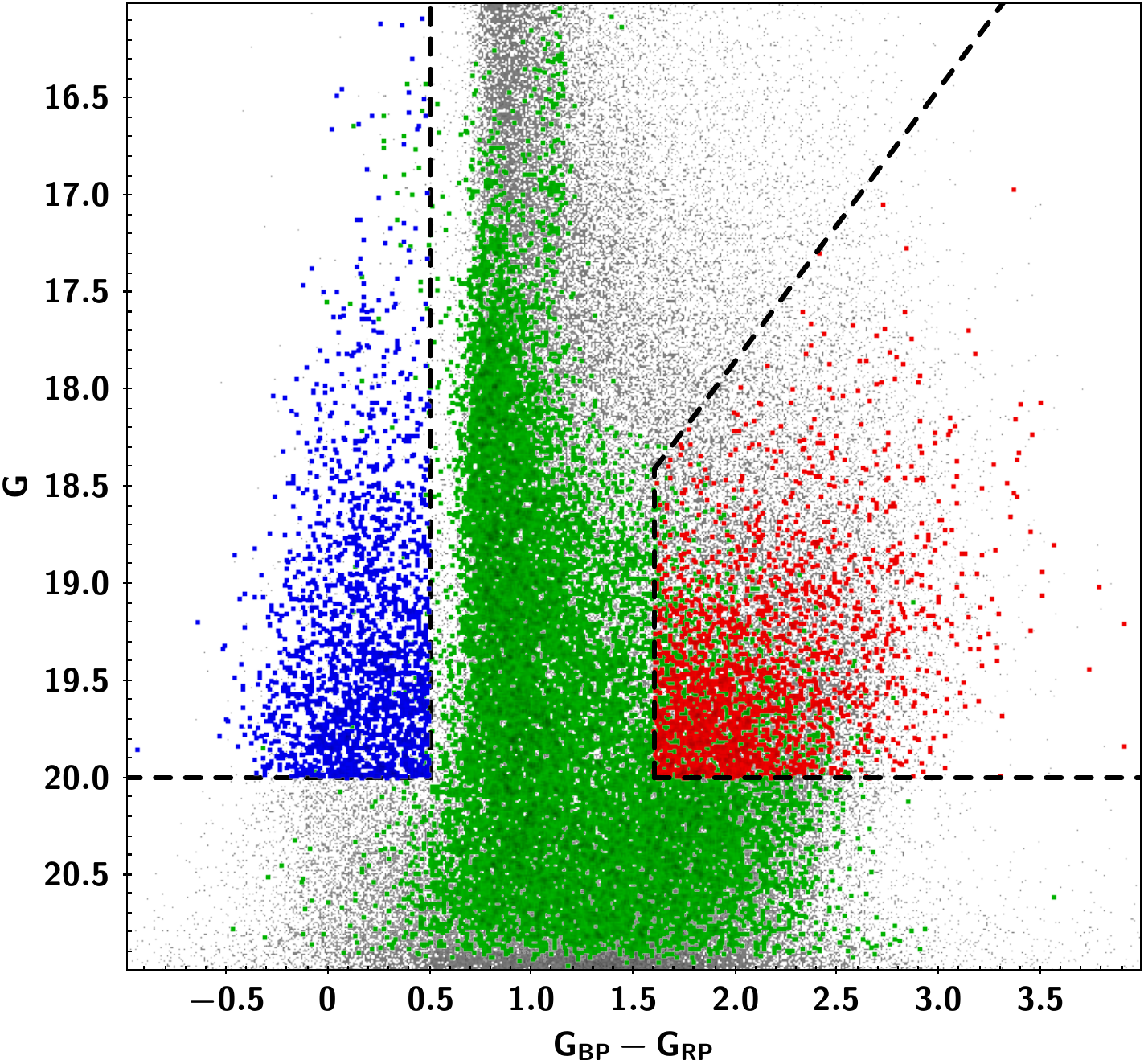}\label{fig:sub_a}}
\end{minipage}
\hfill    
\begin{minipage}{.45\textwidth}
    {\includegraphics[width=\textwidth]{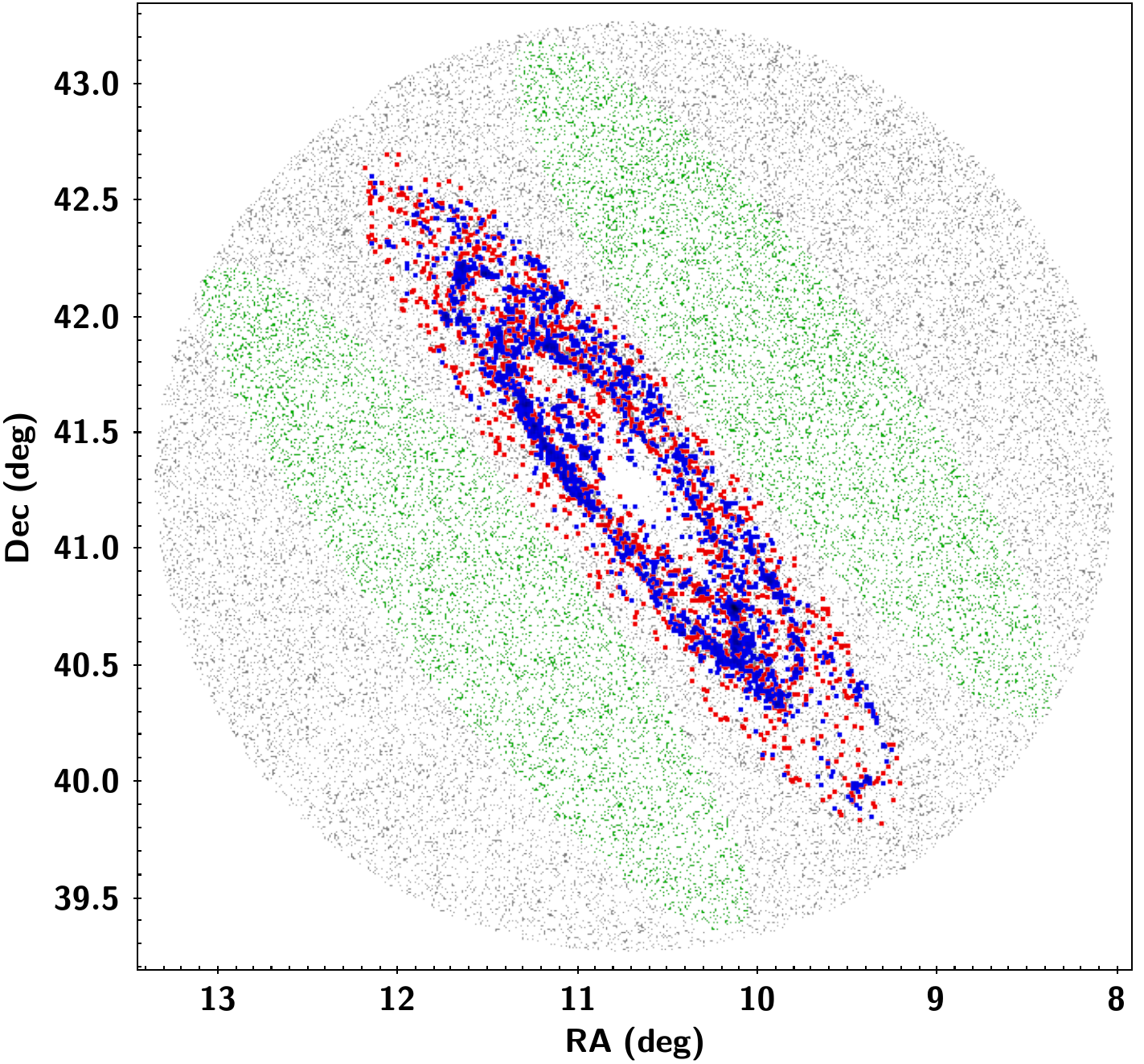}\label{fig:sub_b}}
\end{minipage}
    \caption{M31 sample of resolved stars from Gaia . Left: The color-magnitude diagram for the blue and red supergiants, along with foreground stars in grey. The green stars on the left are from the comparison fields in the figure on the right. Right: The locations of the red and blue supergiants in the disk.
    Figure reproduced from~\cite{2021MNRAS.507.2592S}. }\label{fig:M31gaia}
\end{figure}

 Both {\it Gaia} data release 2 (DR2)~\citep{2019ApJ...872...24V} and early data release 3 (EDR3)~\citep{2021MNRAS.507.2592S} have been used to determine the proper motion of M31. The first step for determining the proper motion involves defining a sample of M31 member stars with accurate astrometry. Due to their bright absolute magnitudes, $G > 16$, blue and red supergiants provide an ideal sample. In addition to being very luminous, blue and red supergiants are well separated from the foreground MW stars and the more numerous main sequence stars in M31. Figure~\ref{fig:M31gaia} shows the {\it Gaia} color-magnitude diagram of the blue and red supergiants used for M31 astrometric measurements, along with their locations in the M31 disk.  

\par Because the member stars are associated with the M31 disk, determining the M31 systemic motion from the {\it Gaia} M31 proper motions requires a model for the rotation velocity of the disk, and this rotational motion must be carefully subtracted out. Performing this subtraction and implementing a maximum likelihood analysis, from EDR3 for the blue supergiant sample the best fitting proper motion is $(\mu_{\alpha \star}, \mu_\delta) = (49 \pm 11,-38 \pm 8)$ $\mu$as/yr~\citep{2021MNRAS.507.2592S}. The proper motion derived from the red supergiant sample is systematically offset from the blue supergiant measurement, possibly due to contamination with the red sample. This measured proper motion is consistent with the combined supergiant sample from DR2 of $(\mu_{\alpha \star}, \mu_\delta) = (65 \pm 18,-57 \pm 15)$ $\mu$as/yr~\citep{2019ApJ...872...24V}. 

\par M31 proper motion measurements from HST and {\it Gaia} are summarized in Table~\ref{tab:M31PMs}. Also shown are the derived relative tangential velocities. Though the DR2 results indicate larger proper motions than those obtained from EDR3, they are consistent within uncertainties. When taking the weighted average of the DR2 results with those from HST, the mean proper motions are more consistent with those obtained from the EDR3 analysis. 

\subsection{Satellite galaxies and outer local group members}
\par Independent of the direct astrometric measurements discussed above, the MW-M31 relative tangential velocity may be determined from indirect measurements. An example of such a method has long been used to determine the MW velocity with respect to the LG barycenter, $\vec v_{MW \rightarrow LG}$, using heliocentric velocities of the OLGMs~\citep{1955AJ.....60..254H,1968Natur.220..868D,1999AJ....118..337C}. The position of the LG barycenter is derived simply assuming that the mass of the LG is dominated by the DM halos of MW and M31. The relative velocity $\vec v_{M31 \rightarrow MW}$ is then derived starting from the measured $\vec v_{MW \rightarrow LG}$, and imposing momentum conservation. 

\par In terms of the MW and M31 velocities with respect to the LG barycenter, the relative velocity may be written as  
\begin{equation}
    {\vec v}_{M31 \rightarrow MW} = {\vec v}_{M31 \rightarrow LG} - {\vec v}_{MW \rightarrow LG}. 
    \label{eq:vm31lgemperical}
\end{equation}
Imposing momentum conservation for the MW-M31 system, the MW and M31 barycentric velocities are related as  ${\vec v}_{M31 \rightarrow LG} = - f {\vec v}_{MW \rightarrow LG}$

where $f = M_{MW}/M_{M31}$ is the ratio of the total mass for the MW to that of M31. Measurements of this ratio from observations are discussed in detail below. In terms of this mass ratio the M31-MW relative velocity is
\begin{equation} 
     {\vec v}_{M31 \rightarrow MW} = - (1+f) {\vec v}_{MW \rightarrow LG}. 
       \label{eq:vm31mwmomentumconservation}
\end{equation} 
This implies that if the MW velocity with respect to the LG barycenter can be independently determined, the M31-MW relative velocity can be determined without a direct measurement of the M31 velocity components. The MW velocity with respect to the LG barycenter may be written in terms of the sum of the MW velocity relative to the Sun, ${\vec v}_{MW \rightarrow \odot}$, and the velocity of the Sun with respect to the LG, ${\vec v}_{\odot \rightarrow LG}$, 
\begin{equation}
{\vec v}_{MW \rightarrow LG} = {\vec v}_{MW \rightarrow \odot} + {\vec v}_{\odot \rightarrow LG}, 
\label{eq:vmwlg}
\end{equation}
with $\vec v_{MW \rightarrow \odot} = - \vec v_{\odot \rightarrow MW}$. 

\par Now assume that there is no net motion in any direction in the OLGM population about any point within the LG. This is a simplified assumption, since it is possible, for example, that the radial infall of satellites into the LG imply a preferential direction of motion. 
With these assumptions, a maximum likelihood procedure may be used to obtain ${\vec v}_{\odot \rightarrow LG}$, solving for ${\vec v}_{\odot \rightarrow LG}$ and the intrinsic velocity dispersion of the OLGM population. Though there are indications that this intrinsic velocity dispersion is a function of position within the LG from samples in simulations~\citep{2022MNRAS.511.6193H}, for simplicity this may be assumed independent of position in the LG. 

\par 
The analysis of~\citet{2014MNRAS.443.1688D} finds $v_{\odot \rightarrow LG} = 312 \pm 25$ km/s and with a dispersion of $58 \pm 11$ km/s, excluding satellite galaxies of the MW and M31. From these measurements, Equation~\ref{eq:vmwlg} may then be used to obtain ${\vec v}_{MW \rightarrow LG}$. Without including the travel velocity term, the Galactic cartesian components are 
\begin{equation}
 {\vec v}_{MW \rightarrow LG}^{(OLGMs)} = (-103.5 \pm 29,  39.8 \pm 26, -60.2 \pm 22) \, {\rm km/s}. 
 \label{eq:vmwlgmeasured} 
\end{equation}
The components of the M31-MW relative velocity may be then obtained from Equation~\ref{eq:vm31mwmomentumconservation}. Assuming that $f = 0.5$, which is consistent with the results discussed below for the mass ratio, the derived tangential velocity components are shown in Table~\ref{tab:M31PMs}. Note that this analysis does not use the heliocentric line-of-sight velocity of M31, which is why the radial velocity component uncertainty is much larger in this case than the direct proper motion case derived above. 

\par The analysis above uses only line-of-sight velocities for the OLGMs, and may be improved with proper motion measurements. Though there are recent measurements of these proper motions~\citep{2022A&A...657A..54B,2024ApJ...971...98B}, the measurement uncertainties are much larger than the expectation for the theoretical dispersion, so it is not possible to extract kinematic information from them. Future {\it Gaia} measurements may be able to improve upon these measurements. 

\subsection{M31 satellites}
\par Now consider a second independent method for measuring the MW-M31 relative tangential velocity. Similar to the OLGM method described above, this is indirect in the sense that it is derived from the kinematic properties of the M31 satellite galaxies, rather than from direct astrometric measurements of M31 member stars. 

\par This method starts by assuming that a population of M31 satellites are dynamically well-mixed and gravitationally-bound to M31. Though the M31 satellites are too faint to have measured proper motions, nearly all of them have measured line-of-sight velocities. Because the lines-of-sight to the different M31 satellites are not perpendicular to the sky plane, the line-of sight velocity contains contributions from both the intrinsic radial velocity component and the intrinsic tangential velocity component. The contributions from the tangential velocity are significant because of the relatively large angular extent of the M31 DM halo. Estimating the M31 virial radius to be 300 kpc, the projected extent of the M31 halo is $\sim 300/770 (180/\pi) \sim 23^\circ$, making this system ideal for determining the proper motion in this manner. This method has also been used to measure the proper motions of MW satellite galaxies~\citep{Kaplinghat:2008sm,2008ApJ...688L..75W}. 

\par The velocity of an M31 satellite may be decomposed as  
\begin{equation} 
{\vec v}_{sat \rightarrow \odot} = {\vec v}_{sat \rightarrow M31} + {\vec v}_{M31 \rightarrow MW} + {\vec v}_{MW \rightarrow \odot}.  
\label{eq:M31satellitevelocity}
\end{equation} 
The first term on the right side is the velocity of a satellite with respect to the center of the M31 DM halo, and the second and third terms have been introduced above. The line-of-sight component of ${\vec v}_{sat \rightarrow \odot}$ is measured for the satellites. The term ${\vec v}_{sat \rightarrow M31}$ contains the intrinsic radial and tangential velocity components of the satellite with respect to M31. This velocity can be extracted under the simple assumption that the satellites are described by an intrinsic velocity dispersion which is constant throughout the halo. With this assumption, Equation~\ref{eq:M31satellitevelocity} can be used to obtain the velocity of M31 with respect to the MW, ${\vec v}_{M31 \rightarrow MW}$. 

\par This method for measuring ${\vec v}_{M31 \rightarrow MW}$ was first exploited in~\cite{vanderMarel:2007yw}. Using an updated population of line-of-sight velocities for a sample of nearly 40 satellites of the M31 subgroup~\citep{2012AJ....144....4M},~\cite{2016MNRAS.456.4432S} further developed this method and updated the M31 proper motion. These authors account for the fact that approximately 15 of the M31 satellites form a thin, planar subgroup of width $\sim 13$ kpc, of which 13 appear to be co-rotating~\citep{2013Natur.493...62I}. This is similar to the plane of satellites which has been identified around the MW~\citep{2005A&A...431..517K}.~\cite{2016MNRAS.456.4432S} test several sub-samples of M31 satellites, examining the influence of the plane of satellites on the results. For the full sample and without an independent constraint on the M31 line-of-sight velocity, the best-fitting tangential velocity is $(v_\alpha,v_\delta) = (34 \pm 70,21 \pm 60)$ km/s. The tangential velocity derived from this method has been shown to depend strongly on the sample of satellites that are selected. 

\par~\cite{2016MNRAS.456.4432S} have also calibrated the method using simulations of LG-like systems, which contain a pair of galaxies with similar kinematic properties to those of the MW and M31. Testing the method for measuring the proper motions on these simulations, the authors find that the extraction of the proper motion is generally unbiased, so that the assumption of a well-mixed system is appropriate. 

\par As compared to the direct proper motion measurements, the uncertainties in the result obtained from M31 satellites are large, and there are several systematics that must be overcome.  For example, it must be conclusively established that the satellites used in the sample are bound to M31. Further, the result is sensitive to identifying which M31 satellites are associated with the plane of satellites. As is the case for the measurement above using OLGMs, proper motions of M31 satellites will ultimately be required to answer this question.

\subsection{M31-MW tangential velocity}
\par
\begin{figure}[h]
\includegraphics[width=3.0in]{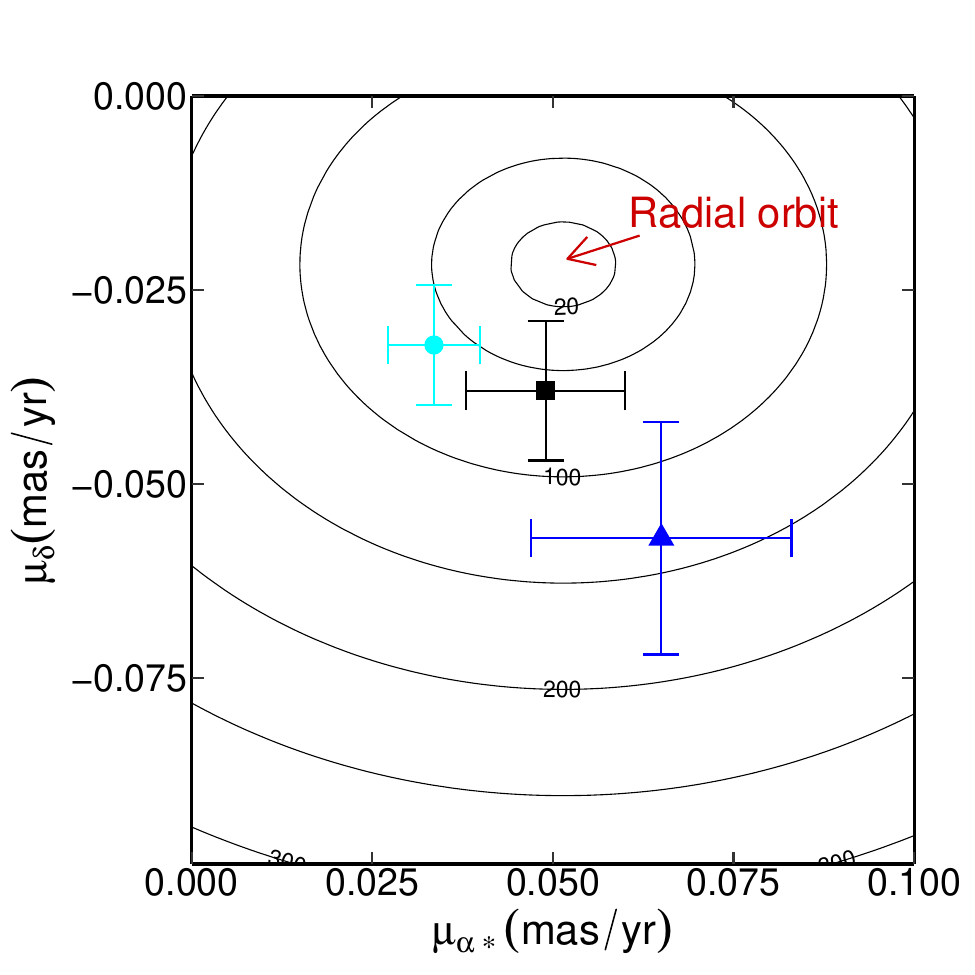}\label{fig:vtcontour}
    \caption{Measurements of the M31 proper motion via the methods discussed in the text, assuming a circular velocity of $239$ km/s. The black square point shows the results from Gaia EDR3, the cyan circle shows the results from HST, and the blue triangle shows results from Gaia DR2. The circular black contours indicate the tangential velocity in km/s. The center point of the circles represents an orbit with zero tangential velocity. The black contours show the radial velocity, without including the possible impact of the LMC.
Abbreviations: DR2, data release 2; EDR3, early data release 3. }\label{fig:M31PM}
\end{figure}

 Figure~\ref{fig:M31PM} summarizes the direct measurements of the MW-M31 proper motion. As is shown there is still a significant scatter in the measurements. It is important to note that the conversion to tangential velocities requires subtraction of $\vec v_{\odot \rightarrow MW}$, and as discussed above there is still some systematic uncertainty in the determination of this velocity. In the corresponding summary of Table~\ref{tab:M31PMs}, the direct measured proper motions from the heliocentric velocity of $-301 \pm 1$ km/s was assumed, which implies a strong bound on $v_{\rm r}$. The Gaia measurements, both with DR2 and EDR3 data, favor a larger value for the tangential velocity as compared to the HST measurement. 

\par Given the heliocentric velocity and for the assumed circular velocity of 239 km/s, a point in the proper motion space ($\mu_{\alpha \star}, \mu_\delta$) corresponds to a purely radial orbit in which the MW and M31 are directly approaching one another. For the above measurements, this proper motion is $(\mu_{\alpha \star}, \mu_\delta) = (0.05,-0.02)$ $\mu$as/yr. This proper motion is in mild tension with the direct Gaia/HST proper motions. For all cases, a purely tangential (circular) orbit is ruled out, so there must be some radial component to the motion. 

\par A relative tangential velocity component implies that there is non-zero orbital angular momentum in the LG. The LG orbital angular momentum per reduced mass for the MW-M31 system is 
\begin{equation}
{\vec L} = {\vec r} \times {\vec v}
\label{eq:Lorb} 
\end{equation}
where $\vec r$ is the relative separation between the MW and M31. Measurements of the LG orbital angular momentum and its implications are discussed in detail below. 

\section{Timing mass}
\par This section outlines the theoretical framework for the timing mass analysis. The section begins with discussion of the theoretical assumptions and how the analysis depends on these assumptions, and then it moves on to obtain the best-fitting LG mass. Systematics that affect the analysis are then considered. 

\subsection{Nature of the orbit} 
\par We start with the simplified assumption that the MW and M31 are two point masses, ignoring their extended structure.  For a two-body orbit, the reduced mass is $\mu = M_{MW} M_{M31}/( M_{MW} + M_{M31})$, and the sum of the masses is $M = M_{MW} + M_{M31}$. Note that here $M$ is assumed to be the total mass of the LG. 
 
\par The MW-M31 two-body system evolves under the influence of the Keplerian gravitational potential 
\begin{equation}
    V_g (r) = - \frac{G \mu M}{r},  
\end{equation}
and the potential that is due to the cosmological constant~\citep{2009A&A...507.1271C}, 
\begin{equation}
    V_\Lambda (r) = - \frac{\Lambda}{6} \mu c^2 r^2, 
\end{equation}
where $\Lambda$ is the cosmological constant, and $r$ is the separation between the two masses. The total potential for the system is $V = V_g + V_\Lambda$. The cosmological constant is related to the observed vacuum energy density, $\Omega_\Lambda$, and the Hubble constant, $H_0$, through $\Lambda = 3 \Omega_\Lambda H_0^2$. More complicated models would account for variation of the vacuum energy from a pure cosmological constant. 

\par
\begin{figure}
\begin{minipage}{.45\textwidth}
    {\includegraphics[width=\textwidth]{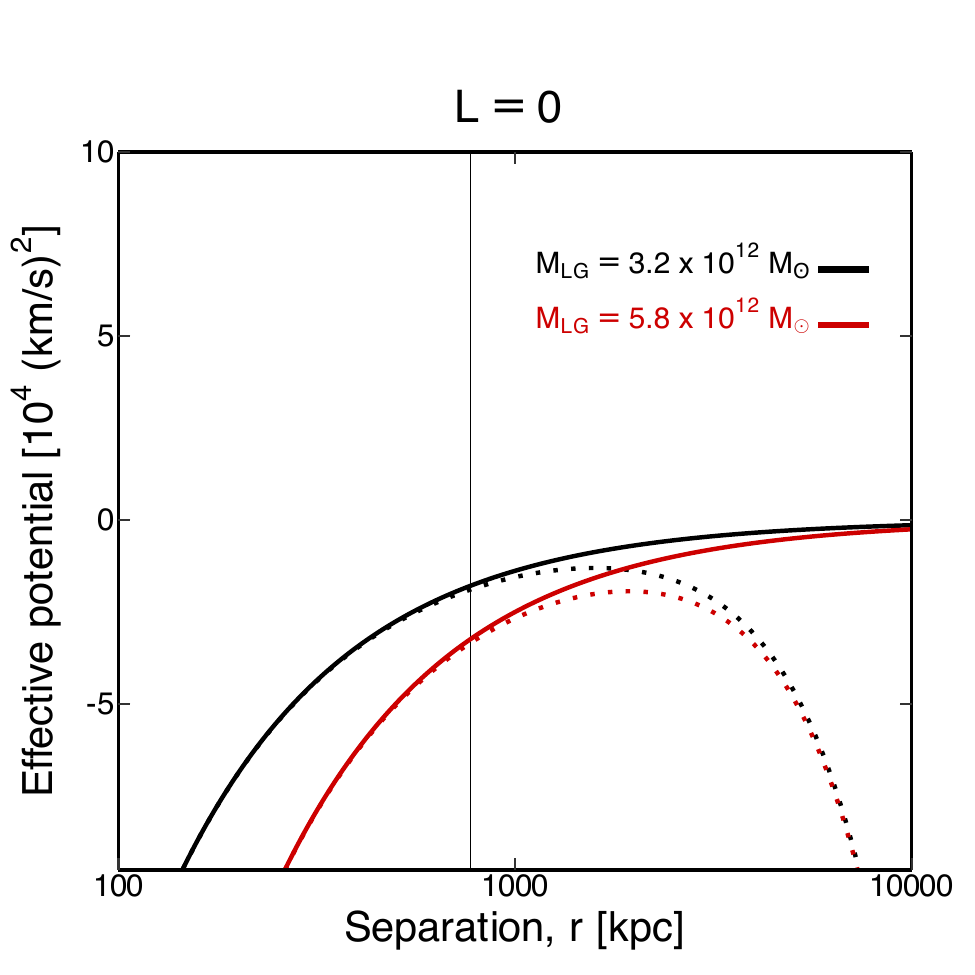}\label{fig:eff_a}}
\end{minipage}
\hfill    
\begin{minipage}{.45\textwidth}
    {\includegraphics[width=\textwidth]{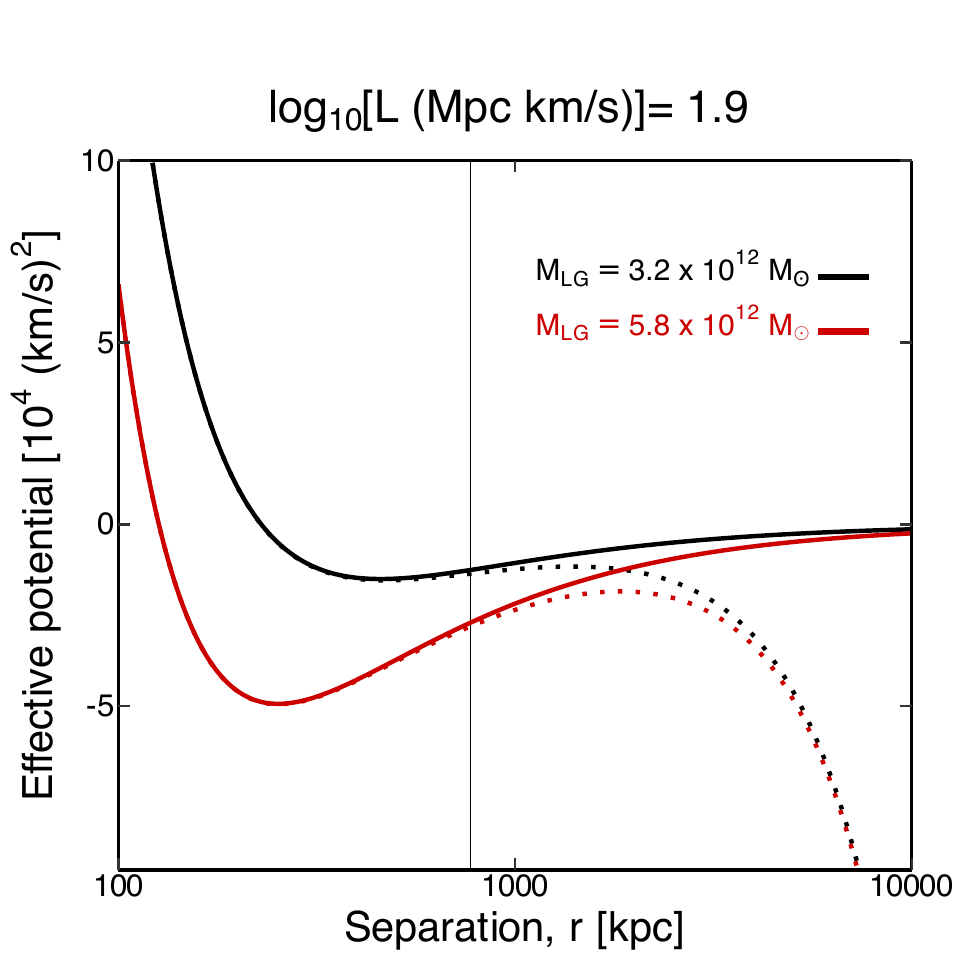}\label{fig:eff_b}}
\end{minipage}
    \caption{Effective potential versus separation for the MW-M31 two-body orbit. The solid curves show the case of a pure Keplerian model, while the dotted curves show the case in which the cosmological constant is included, with cosmological parameters given in the text. (a) The case of zero orbital angular momentum. (b) The case with an orbital angular momentum $\log_{10} [ L ({{\rm Mpc} \, \rm km/s}) ] = 1.9$, which corresponds to best-fitting values for these parameters. Different assumptions for the LG mass are indicated. In the right panel, the first local minimum shows the radius of a stable circular orbit, while the local maximum in the dotted curve shows the radius of the unstable circular orbit. The vertical line shows the present day separation between the MW and M31.}\label{fig:effectivepotential}
\end{figure}

 The dynamics of the system is described by the Lagrangian per reduced mass, so the potential may be re-scaled as $V \rightarrow V/\mu$. The effective potential per reduced mass is then  
\begin{equation} 
V_{\rm eff} = - \frac{GM}{r} + \frac{L^2}{2 r^2} - \frac{1}{2} \Omega_\Lambda H_0^2 r^2
\label{eq:Veff} 
\end{equation} 
where $L$ is the magnitude of the orbital angular momentum from Equation~\ref{eq:Lorb}. The total energy per reduced mass follows as $E = v_{\rm r}^2 / 2 + V_{\rm eff}$, where $v_{\rm r}$ is the radial component of the relative velocity. The second term on the right-hand side of Equation~\ref{eq:Veff} is the angular momentum barrier, which increases the potential at small radii and establishes a minimum separation. Accounting for just the Keplerian potential and the angular momentum barrier, and examining the extrema of $V_{\rm eff}$, there is a local minimum that depends on $M$ and $L$, which implies that the system admits a stable, circular orbit with a radius, $r_{stab}$ (see Figure~\ref{fig:effectivepotential}). In this case, orbits are bound with a finite maximal separation, or apocenter, for $E < 0$, and unbound with an infinite maximal separation for $E \ge 0$. 

\par The third term on the right-hand side of Equation~\ref{eq:Veff} arises from the cosmological constant, for which the potential scales as $r^2$. Examining the extrema of $V_{\rm eff}$ including this term, the presence of the cosmological constant term adds a local maximum to the effective potential, and a corresponding radius of an unstable circular orbit, $r_{unstab}$, with a radius that depends on the combination $\Omega_\Lambda H_0^2$, as well as $L$ and $M$ (See Figure~\ref{fig:effectivepotential}).  This radius is always outside the stable circular orbit radius which results from the Keplerian potential, $r_{unstab} > r_{stab}$. For a given $L$ and $M$, there is a critical radius at which $r_{unstab} = r_{stab}$, corresponding to circular orbits of marginal stability. In the region of $L$ and $M$ space in which $V_{\rm eff}$ monotonically decreases as a function of $r$, no circular orbits are admitted in the system. 

\par Note that energy and angular momentum are conserved when including the cosmological constant. However, the criteria for classifying orbits as bound or unbound is more complicated as compared to the case of a pure Keplerian potential. In particular, with the cosmological constant unbound orbits with $E < 0$ are admitted. This is not possible in the Keplerian case, for which all orbits with $E < 0$ are bound. As in the case of the Keplerian potential, with the cosmological constant orbits with $E \ge 0$ are unbound. 

\par The evolution of the separation between the two galaxies is
\begin{equation} 
\ddot r = - \frac{GM}{r^2} + \frac{L^2}{r^3}  + \Omega_\Lambda H_0^2 r 
\label{eq:equationofmotion}
\end{equation} 
with the dots denoting derivatives with respect to time, $t$. The Keplerian potential induces a force directed radially inward, while the cosmological constant term induces a force directed radially outward. In the limit that the cosmological constant term is negligible, the solution for $r(t)$ may be written in terms of Keplerian orbital elements
\begin{eqnarray}
r &=& a (1- e \cos \eta) \\
t &=& (a^3/GM)^{1/2} (\eta - e \sin \eta) 
\label{eq:rddot} 
\end{eqnarray}
where $a$ is the semi-major axis, $\eta$ is the eccentric anomaly, and $e$ is the eccentricity. At the pericentric passage, the relative separation is $r_{\rm p} = a(1-e)$, and $\eta = 0$, while at apocentric passage, the relative separation is $r_{\rm a} = a(1+e)$, and $\eta = \pi$. The radial velocity and the tangential velocity may be written in terms of the Keplerian orbital elements as 
\begin{eqnarray}
v_{\rm r} &=& \frac{GM}{r} \frac{e \sin \eta}{1 - e \cos \eta}
\label{eq:vrkeplerian} 
\\
v_{\rm t} &=& \frac{GM}{r} \frac{\sqrt{1 - e^2}}{1 - e \cos \eta}. 
\label{eq:vtkeplerian} 
\end{eqnarray}
In the limit of a pure radial orbit, $e \rightarrow 1$ and $v_{\rm t} = 0$, which implies that $L \rightarrow 0$, so the angular momentum barrier is negligible and highly eccentric, elliptical orbits with $r \rightarrow 0$ are admitted. As $L$ is increased, for a fixed mass the minimum separation that the bodies can achieve increases. 

\par For the LG, the Keplerian term dominates the contribution to the potential. The impact of the cosmological constant may be estimated by comparing the Keplerian period of the system, $T_{kep} = 2 \pi \sqrt{ a^3/GM }$, to the period related to the cosmological constant, $T_\Lambda = 2\pi/c\sqrt{\Lambda} = 2\pi/(c \sqrt{3 \Omega_\Lambda} H_0)$. For the measured cosmological parameters, $\Omega_\Lambda = 0.7$ and $H_0 = 70$ km s$^{-1}$ Mpc$^{-1}$, the period is $T_\Lambda = 61$ Gyr. Then defining the ratio of the periods as $\lambda = (T_{kep}/T_\lambda)^2$, and working in the limit of a perturbed Keplerian orbit, using the expansion parameter $\lambda < 1$, a new set of equations corresponding to Equations~\ref{eq:vrkeplerian} and~\ref{eq:vtkeplerian} may be derived~\citep{Benisty:2023vbz}. 

\par The formulae above present a system of equations in which $M, \eta, a$, and  $e$ may be solved for given the observed values of $r, t, v_{\rm r}, v_{\rm t}$. While they are strictly analytic in the matter-dominated case, for vacuum energy they are an approximation that relies on the accuracy of the perturbation theory. Including the cosmological constant in the analysis mildly shifts the fitted values of the parameters describing the orbit. 
 
\par The most general solutions to Equation~\ref{eq:equationofmotion} may be obtained numerically, starting from initial conditions at the present epoch, evolving the orbit backwards in time. Since in a two-body orbit the angular momentum is conserved, there are four parameters that must be specified for the initial conditions, namely the present-day values of $r$, $v_{\rm r}, v_{\rm t}$, and the angular position.  

\subsection{Orbit from observed Local Group properties} 
\par From the orbit equation derived above, several scenarios can be considered for the MW-M31 dynamics, while only imposing the assumption that they are on first approach. First consider the scenario in which the orbit is purely radial, with $v_{\rm t} = 0$. This is the original motivation of the timing analysis, in which the MW and M31 were in close proximity in the distant past. The timing mass is obtained by solving for the mass with the additional assumption that $r \rightarrow 0$ as $t \rightarrow 0$, which corresponds to a lookback time of $13.8$ Gyr. 
Taking $v_{\rm r} = -109 \, {\rm km/s}$, the LG timing mass is 
\begin{equation}
M_{TM}^{(v_{\rm t}=0)} = 4.2 \times 10^{12} \, {\rm M}_\odot.  
\label{eq:timingradial}
\end{equation}

\par A non-zero present-day tangential velocity changes the deduced LG mass. Assuming angular momentum is conserved, if the MW and M31 had a non-zero tangential velocity at their formation, they have a non-zero pericenter and there is a minimum separation the two halos achieved in the distant past. The timing analysis can be implemented as above under this assumption, with the separation minimized as $t \rightarrow 0$. For the EDR3 measurement with its uncertainties, and the same radial velocity assumption as above, 
\begin{equation}
M_{TM}^{(EDR3)} = (5.8 \pm 0.7) \times 10^{12} \, {\rm M}_\odot. 
\label{eq:timingedr3}
\end{equation}
Comparison of these results shows that the presence of a tangential velocity increases the LG timing mass. For the same radial velocities but taking the lower tangential velocities consistent with the HST data, the timing mass reduces to $4.8 \times 10^{12} {\rm M}_\odot$. 

\par
\begin{figure}[h]
\includegraphics[width=3.0in]{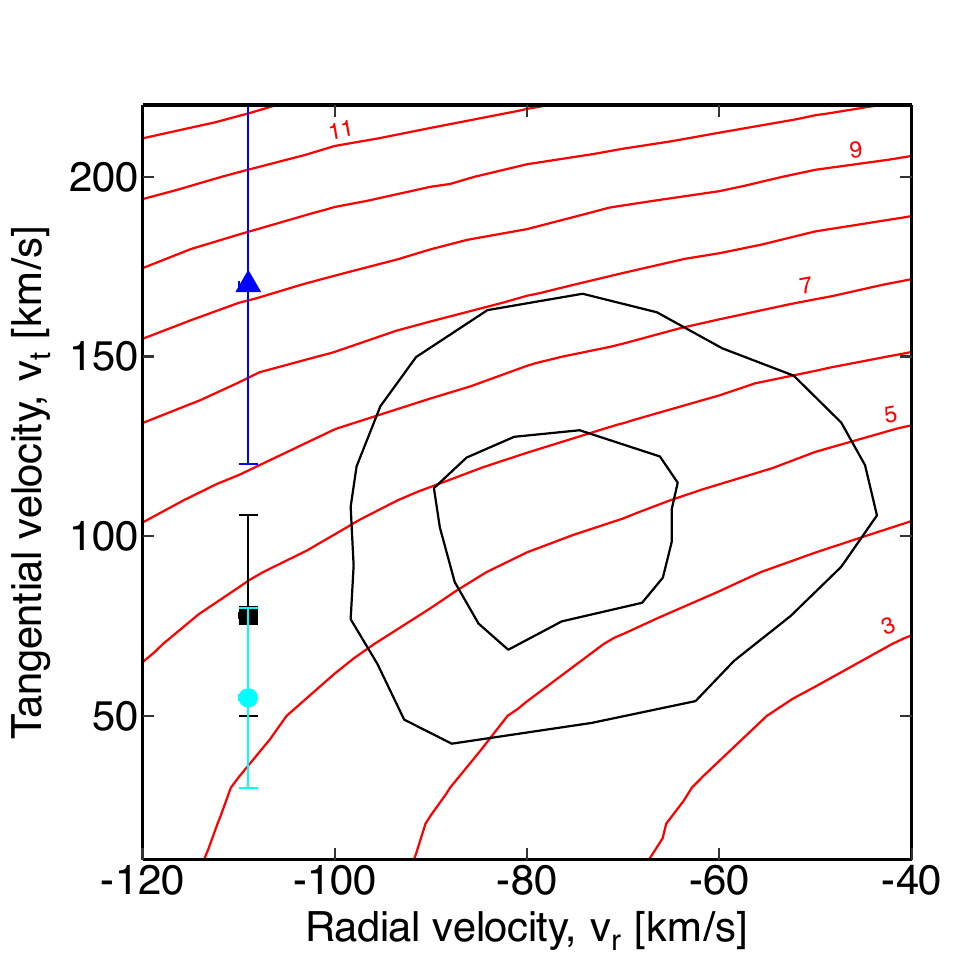}
\caption{Contours of the timing mass in the plane of the MW-M31 relative radial and tangential velocity. Red contours are the timing mass in units of $10^{12}$ M$_\odot$. Data points are measurements of the tangential and radial velocity as derived from the direct measurements of the M31 proper motion from Table~\ref{tab:M31PMs}: black square point shows the results from Gaia EDR3, the cyan circle shows the results from HST, and the blue triangle from Gaia DR2. The black contours estimate the impact of the LMC~\citep{2022ApJ...928L...5B}, shifting the EDR3 measurement to a region of $v_{\rm r}, v_{\rm t}$ space.  
Abbreviations: DR2, data release 2; EDR3, early data release 3.
}
\label{fig:vrvtmass}
\end{figure}

\begin{figure}
\begin{minipage}{.42\textwidth}
    {\includegraphics[width=\textwidth]{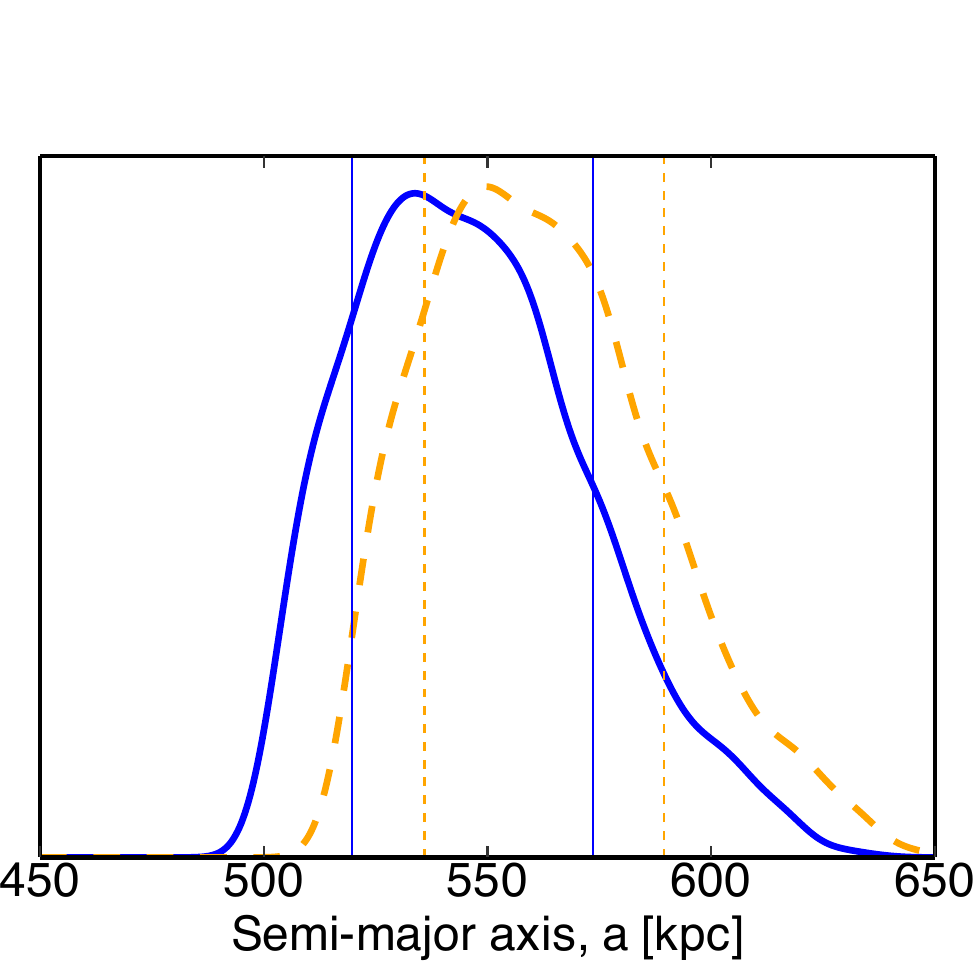}\label{fig:pa}}
\end{minipage}
\hfill    
\begin{minipage}{.42\textwidth}
    {\includegraphics[width=\textwidth]{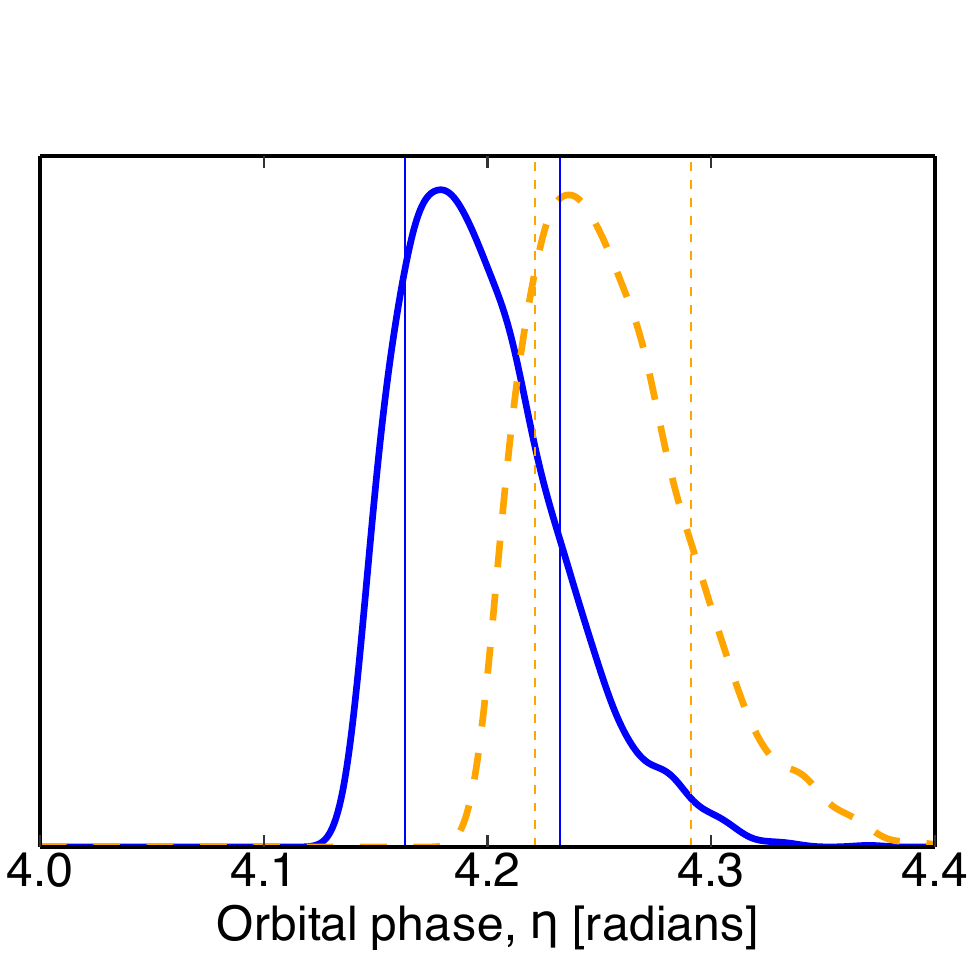}\label{fig:peta}}
\end{minipage}
\medskip 
\begin{minipage}{.42\textwidth}
    {\includegraphics[width=\textwidth]{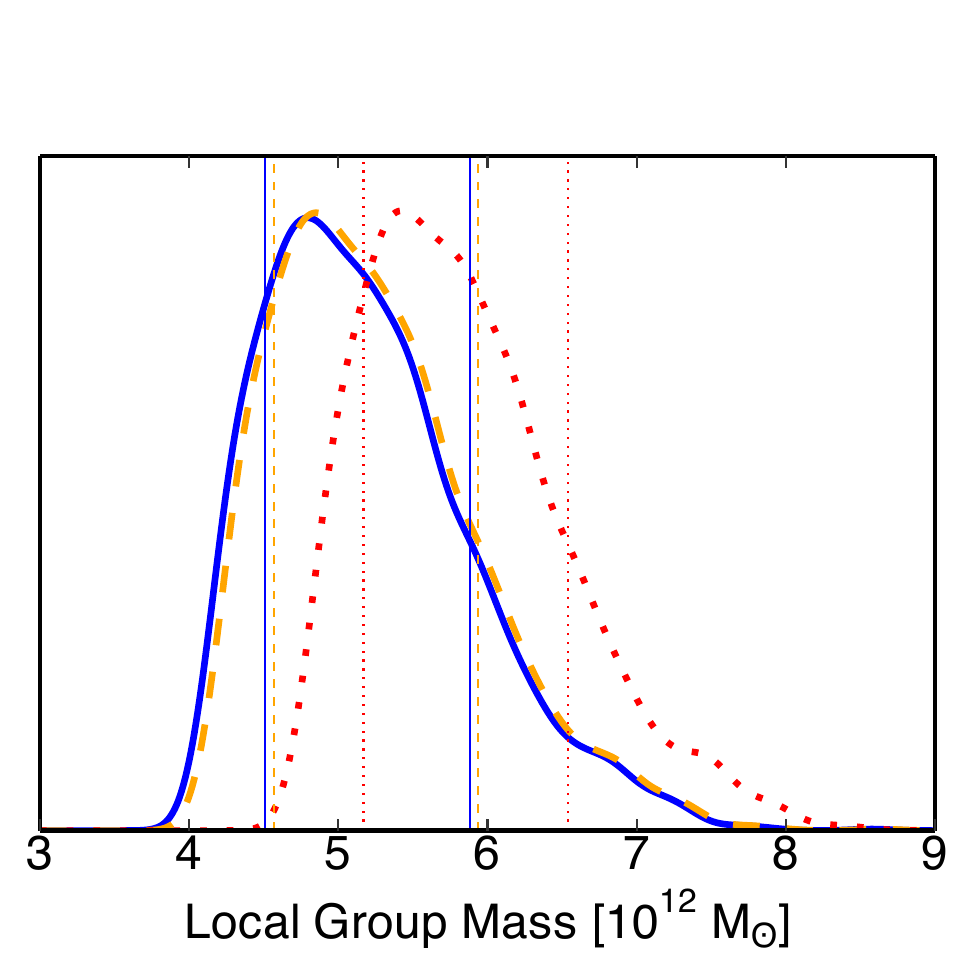}\label{fig:pmass}}
\end{minipage}
\hfill
\begin{minipage}{.42\textwidth}
    {\includegraphics[width=\textwidth]{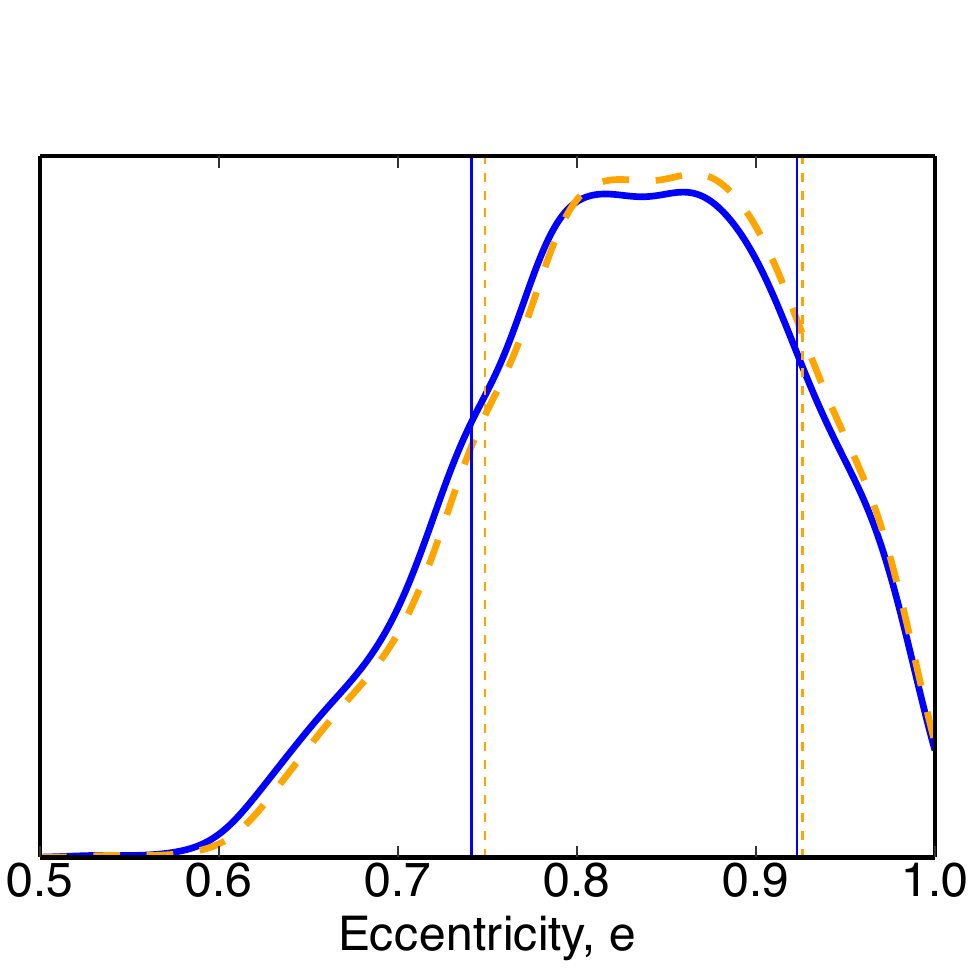}\label{fig:pecc}}
\end{minipage}
   \caption{MW-M31 orbital parameters for different model assumptions. Orange histograms show the results from the analytic analysis without the cosmological constant, and the blue histograms include the cosmological constant using an analytic formalism~\citep{Benisty:2023vbz}. For the mass, the red-dotted histogram results from the full analysis of the equation of motion including the cosmological constant. Vertical lines show the 68\% containment regions for the respective distributions.}\label{fig:TAparameters}    
\end{figure}

 Most generally, Figure~\ref{fig:vrvtmass} shows the deduced timing mass for a range of the relative radial and tangential velocities. Large values of the tangential velocity, which are consistent with the indirect proper motion measurements described above, imply an LG mass approaching $\sim 10^{13}$ M$_\odot$, which is nearly an order of magnitude larger than the sum of the MW and M31 masses, as discussed below. For the EDR3 tangential velocity and range of uncertainties, the best fitting parameters that describe the MW-M31 orbit may be determined. The best fitting values for the semi-major axis, eccentric anomaly, and eccentricity are, respectively, $a = 562 (546)$ kpc, $\eta = 4.3 (4.2)$, and $e = 0.84 (0.83)$, with the values inside the parenthesis including the cosmological constant~\citep{Benisty:2023vbz}, while those outside the parenthesis exclude it. The uncertainties are generated from monte carlo sampling of the uncertainties on the tangential velocity. Figure~\ref{fig:TAparameters} shows the full distributions for each of these parameters, including the implied LG mass.

\par Generally, the timing analysis is sensitive to any effect that changes the MW-M31 relative velocities, which is captured in Figure~\ref{fig:vrvtmass}. In addition to the shift due to the impact of the LMC, which is modeled in detail below, a smaller shift arises from changing the LSR velocity~\citep{2014MNRAS.443.2204P}. It is also interesting to note that the presence of a massive LMC also shifts the barycenter of the LG, relative to the case in which only the MW and M31 are the dominant components.~\cite{2016MNRAS.456L..54P} model this barycenter shift and show that the timing mass reduces by approximately 20\% over the range of plausible masses for the LMC. 

\par Though it provides a reasonable estimate for the mass of the LG, the timing mass calculation is clearly a simplification. There are several physical effects which are not captured in the orbit equations presented above. Examples of these effects may include a significant amount of substructure in the LG, the time dependence of LG mass, and possible interactions with nearby galaxies. One means to address the systematics are by appealing to simulations, which are considered in the detail below.

\section{Masses of Local Group constituents}

\par As emphasized the timing mass method is unique in that it provides an estimate of the total LG mass, under the assumption that the MW and M31 comprise the entire mass of the LG. How does the timing mass compare to measurements of the sum of the individual masses for the MW and M31? Further, how does it compare to LG mass measurements from different methods? 
  
\par Though a substantial amount of literature has been devoted to measuring the MW and M31 DM masses, there remains scatter and systematic uncertainties that must still be understood. Of particular importance is determining the total radial extent of the halos, and measuring the mass contained within this outer radius. Typically, the mass is best estimated within an interior radius associated with the characteristic radius of a tracer population, and extrapolation to a larger radius is performed to estimate the total halo mass. So estimating the mass depends strongly on the radius within which the mass is measured, and the population of objects used to obtain the measurement. As an additional complication, halos are not perfectly spherical systems with well-defined boundaries, and contain both baryonic and DM substructure. 

\par This section discusses mass measurements for the primary LG constituents, the MW and M31, as well as measurements for the larger dwarf galaxies in the LG. These measurements are compared to the timing mass determined above. To frame the discussion, it is important to establish a uniform definition of DM halo mass-- this section begins by establishing such a definition.

\subsection{Definition of dark matter halo mass}

\par DM halos form through a process of smooth accretion and hierarchical merging of lower-mass halos. Simulations find that halos are well-described by an Einasto or Navarro-Frenk-White (NFW) density profile~\citep{2010MNRAS.402...21N}, the latter of which is defined through the relation $\rho(r) = \rho_s/[(r/r_s) (1 + r/r_s)^2]$, with the scale density $\rho_s$ and the scale radius $r_s$ as the two parameters. The baryonic content of a halo modifies the DM density. Because halos are continually evolving through accretion and mergers, there is no natural definition of the halo outer boundary, and therefore no straightforward definition for the total halo mass. A reasonable, approximate definition is the virial mass, $M_{vir}$, which is derived from the spherical collapse model that relates the halo virial radius, $r_{vir}$, to cosmological parameters~\citep{1998ApJ...495...80B}. Integrating the spherical density distribution, for example from an NFW model, out to the virial radius provides an estimate for the virial mass.

\par In place of the scale density and scale radius, DM halos may be characterized by their concentration and characteristic radius. The concentration of a halo is defined as the ratio of the virial radius to the scale radius, $c = r_{vir}/r_s$. Simulations find that the larger mass halos have the smallest concentrations, and specifically for halo masses $> 10^{12}$ M$_\odot$ the concentration is nearly constant as a function of mass. For lower mass halos, the concentrations gradually rise~\citep{2014MNRAS.441.3359D}. It is also observed that for lower mass halos there is more evolution of the halos in redshift~\citep{2011ApJ...740..102K}. Using a sequence of zoom-in simulations, the concentration-mass relation extrapolates to halos of mass as low as $10^{-13}$ M$_\odot$~\citep{2020Natur.585...39W}.  

\par As an alternative to the virial mass, a second standard mass definition for a halo is $M_{200}$. This is defined as the mass within a radius at which the density is 200 times the critical density of the Universe, where $r_{200}$ is the radius at which this density is achieved. There is a similar definition of concentration in terms of this mass, $c_{200} = r_{200}/r_s$. For MW and M31 mass halos, $c \sim 10$ and $c_{200} \sim 7.4$, respectively, which leads to $M_{vir}/M_{200} \sim 1.2$. 

\par Halo density models or concentration-mass relations are of practical importance because they can be used to relate halo mass within a given radius to the total halo virial mass. As discussed below, as applied to kinematic data, it is common to estimate the halo mass at a radius interior to the virial radius, since this is where the majority of the halo tracers reside. A measurement at an interior radius can then be related to the halo virial radius, given input on the structure of halos from simulations.

\subsection{Milky Way mass} 

\par The mass and extent of the MW halo is measured using the kinematics of tracer populations within the halo. The tracer populations that have been used include stars, dwarf satellite galaxies, and globular clusters. Ideally, such a population extends to the edge of the MW halo and is sampled with a sufficient density. However, since the  density naturally declines towards the outer edge of the halo, obtaining a precise and unbiased estimate for the total halo mass is subject to incompleteness of the tracer population. 

\par Given a tracer population, the total MW halo mass has been measured through either distribution-function-based or moment-based methods. Both methods assume that the tracer population is in dynamical equilibrium. The distribution-function-based method starts by assuming a form for the density of tracers per unit volume of phase space, $f$. While these methods are the most theoretically reliable, the accuracy depends on the assumed form of $f$, which is difficult to predict from theoretical first principles. 

\par Similar to distribution-function based methods, moment-based methods assume that the population is in dynamical equilibrium. However, these methods do not require a specific assumption for $f$, since the observations are equivalent to density or velocity moments of $f$. In moment-based methods, the velocity dispersions are binned as a function of radius, and then the velocity dispersions are fit to equilibrium models through the jeans equation. While moment-based methods are in principle more general than distribution-function based methods, information is lost in extracting the moments. Both distribution-function based and moment-based methods require an assumption for the anisotropy of the velocity distribution, which is defined as the ratio of the dispersions in the tangential and the radial direction.

\par There have been several recent applications of the above mass estimator methods to obtain the MW virial mass.~\cite{2020SCPMA..6309801W} review recent measurements of the MW mass, here  recent results are highlighted that are most sensitive to the extended DM halo.~\cite{2019ApJ...873..118W} apply mass estimators previously developed for MW halo globular clusters, and obtain $M_{vir}={1.28}_{-0.48}^{+0.97} \times {10}^{12}$ M$_\odot$.~\cite{2019MNRAS.484.5453C} develop an estimate of the MW mass based on energy and angular momentum distributions derived from simulations. Using kinematics of MW satellites, they find a mass of $1.17_{-0.15}^{+0.21} \times 10^{12}$ $M_\odot$.~\cite{2020MNRAS.494.4291C} use adiabatic contraction models of the MW halo motivated from simulations, and use this to fit to the Gaia DR2 rotation curve. They find a best-fitting MW mass of $1.08_{-0.14}^{+0.20} \times 10^{12}$ $M_\odot$.~\cite{2020MNRAS.498.5574E} find that the MW mass is biased high without accounting for the LMC, and obtain a MW mass of $10^{12}$ M$_\odot$. Modeling the orbit of the outer MW satellite Leo I, and assuming that this satellite is bound to the MW,~\citet{2013ApJ...768..140B}  estimate the MW mass to be $1.6 \times 10^{12}$ M$_\odot$. The resulting weighted mean of these estimates for the MW halo mass using tracers at large distance is
\begin{equation}
M_{vir}^{(MW)} = (1.2 \pm 0.5) \times 10^{12} \, {\rm M}_\odot. 
\end{equation} 

\par Additional measurements of the MW halo mass come from the analysis of the local escape velocity~\citep{2014A&A...562A..91P}. This measurement requires defining a sample of stars with the largest Galactocentric velocity, then fitting this sample to an empirical shape for the velocity distribution near the escape velocity.~\cite{2019MNRAS.485.3514D} model the local distribution of the fastest stars from simulations, deducing a MW mass of $M_{200} = 1.00 \times 10^{12} M_\odot$. Using radial velocities and proper motions from the {\it Gaia} satellite,~\citet{Roche:2024gcl} measure the escape velocity over a range of Galactocentric radii of $\sim 4-10$ kpc. They find a best-fitting mass of $M_{200} = 0.55 \times 10^{12}$ M$_\odot$, which is on the lower end of MW-mass estimates. 

\subsection{M31 mass} 

\par Modern estimates for the mass of the extended DM halo of M31 have been obtained using tracer populations similar to those obtained for the MW. Also similarly, the challenge is transforming from enclosed mass to total M31 halo virial mass. However, given the larger distance to the M31 tracer populations, the kinematics of these populations are not as well measured, resulting in fewer and less accurate mass estimates. 

\par~\cite{2000MNRAS.316..929E} develop a distribution-function-based method, in combination with an analytical model for the density of the DM halo that scales as $r^{-5}$ at large radii and therefore has a finite total mass. Using kinematics of satellites, globular clusters, and planetary nebulae, they estimate the total virial mass of M31 to be $12.3_{-6}^{+18} \times 10^{11}$ M$_\odot$. Using a jeans-based method,~\cite{2010MNRAS.406..264W} develop a mass estimator, assuming that the potential, the tracer population density profile, and the velocity anisotropy profile obey power laws over the radial range of their measurement. They find the M31 mass within 300 kpc to be $(1.4 \pm 0.4) \times 10^{12}$ $M_\odot$. Given the nature of this estimator the result depends mildly on the assumed velocity anisotropy.~\citet{2012ApJ...752...45T} use the velocities of satellites to obtain a virial mass of $1.2_{-0.7}^{+0.9} \times 10^{12}$ M$_\odot$.~\citet{2014ApJ...789...62H} extended, finding evidence for a prolate M31 halo and a mass of $1.82_{-0.39}^{+0.49} \times 10^{12}$ M$_\odot$ within 200 kpc for an assumed NFW profile. Modeling the rotation curve,~\citet{2024MNRAS.528.2653Z} find a M31 virial mass of $1.14 \times 10^{12}$ M$_\odot$. Stellar streams may also be used to constrain the mass of M31~\citep{2004MNRAS.351..117I,2023ApJ...944....1D};~\cite{2013MNRAS.434.2779F} use the observed giant southern stream to measure the mass of M31 to be $2 \times 10^{12}$ M$_\odot$. All recent M31 mass estimates are reviewed in~\citet{2023arXiv230503293B}. 

\par Considering just the mass estimators that use tracers of the M31 DM halo at large distances, the weighted mean of the M31 DM halo mass is 
\begin{equation} 
M_{vir}^{(M31)} =  (1.3 \pm 0.6) \times 10^{12} \, {\rm M}_\odot. 
\end{equation} 
Though there is some scatter in the individual mass measurement results, modern results generally point to M31 being the most massive component of the LG. This is consistent with results deduced from simulations, as described below. In contrast to the MW, there is evidence that M31 has had more recent merger activity~\citep{2018NatAs...2..737D}. This fact complicates the equilibrium modeling using the tracer populations above, and presents an additional systematic uncertainty that is not present for the MW, which appears to have had a quieter recent merger history. 

\par The escape velocity of M31 has also been used to constrain its mass.~\citet{2018MNRAS.475.4043K} determine the escape velocity of M31 to be $\sim 470$ km s$^{-1}$ at a galactocentric distance of 15 kpc, and estimate the virial mass and radius of the galaxy to be $0.8 \times 10^{12}$ M$_\odot$. Similar to the case of the MW, this is on the lower end of the DM halo mass measurements. 

\subsection{Masses of Local Group dwarf galaxies} 
\par Mass measurements of the LG dwarf galaxies may be used to determine their contribution to the total LG mass budget. For the most massive dwarf galaxies in the LG, the DM distribution may be estimated from the stellar and gas rotation curves. As in the case of the MW and M31 mass estimates, converting from enclosed mass to total halo mass requires appealing to the halo concentration mass relation, and also where appropriate accounting for tidal stripping of the DM halos.  

\par The spiral galaxy M33 is likely the third most massive galaxy in the LG. From estimates of its orbit it is unclear whether M33 is a bound satellite of M31~\citep{2020MNRAS.493.5636T} or has experienced a tidal interaction with M31~\citep{2024A&A...685A..38C}.~\cite{LopezFune:2016uun} performed a detailed analysis of the baryonic and DM contributions to the M33 rotation curve. These authors find that the central density is consistent with both a cored and cusped model, and for a cusped NFW profile they find a total halo virial mass of $5 \times 10^{11}$ M$_\odot$ and a halo concentration of $c = 9.2$, which is consistent with the concentration mass relation predicted from simulations. 

\par The kinematics of the LMC have been studied by many authors over the years. The rotation curve is now measured with great precision with {\it Gaia}~\citep{2016ApJ...832L..23V}. From HST proper motions, ~\cite{2014ApJ...781..121V} measure a mass of the LMC within 8.7 kpc of $1.7 \times 10^{10}$ M$_\odot$.  Independent measurements of the LMC mass come from measurement of the perturbation from the Orphan Stream~\citep{2019MNRAS.487.2685E} and the streams in S5~\citep{2021ApJ...923..149S}.~\cite{2024arXiv240114458W} consider the mass estimators previously applied to the MW and M31, using LMC globular clusters to obtain a mass within 13.7 kpc of $2.7 \times 10^{10}$ M$_\odot$, which corresponds to a virial mass of $1.8 \times 10^{11}$ M$_\odot$.  

\par The remaining lower luminosity dwarf galaxies appear to constitute a small component of the LG mass budget. For example, using an axisymmetric jeans model for its kinematics and photometry,~\cite{2021MNRAS.500..410L} find a virial mass for the Wolf-Lundmark-Mellotte galaxy of $\sim 2 \times 10^{10}$ M$_\odot$.~\cite{2003MNRAS.340...12W} find a halo mass of $3 \times 10^9$ M$_\odot$ for NGC 6822.~\cite{2010ApJ...711..361G} construct two-integral models of NGC 147 and NGC 185, and find that in both cases a significant DM component is required, with minimum DM masses of $\sim 6 \times 10^8$ M$_\odot$ for NGC 147 and $\sim 7 \times 10^8$ M$_\odot$ for NGC 185. From direct measurements of their stellar kinematics, dwarf spheroidal galaxies have characteristic DM masses lower than these measurements~\citep{2022NatAs...6..659B}. 

\subsection{Summary: Local Group mass estimates}

\par The above results imply that approximately $\sim 7 \times 10^{11}$ M$_\odot$ is contained in the two most massive LG dwarf galaxies. Even taking the higher end of their respective allowed mass range, the sum of the MW, M31 and dwarf galaxy masses is still less than the timing mass estimated in Equation~\ref{eq:timingedr3}. How should this difference be interpreted? On the one hand, the MW and M31 results may be systematically low, which would imply that there is approximately a factor of two times more mass in the extended halo than deduced from the observed kinematics. On the other hand, the difference may indicate that there is a systematic uncertainty in the timing mass calculation. Possibilities include large uncertainties in the radial and tangential velocities, or a bias in the estimate that results from theoretical assumptions. This latter issue is discussed in detail in the simulation section below. 

\par
\begin{figure*}[!h]
\centering 
    \includegraphics[width=3.0in]{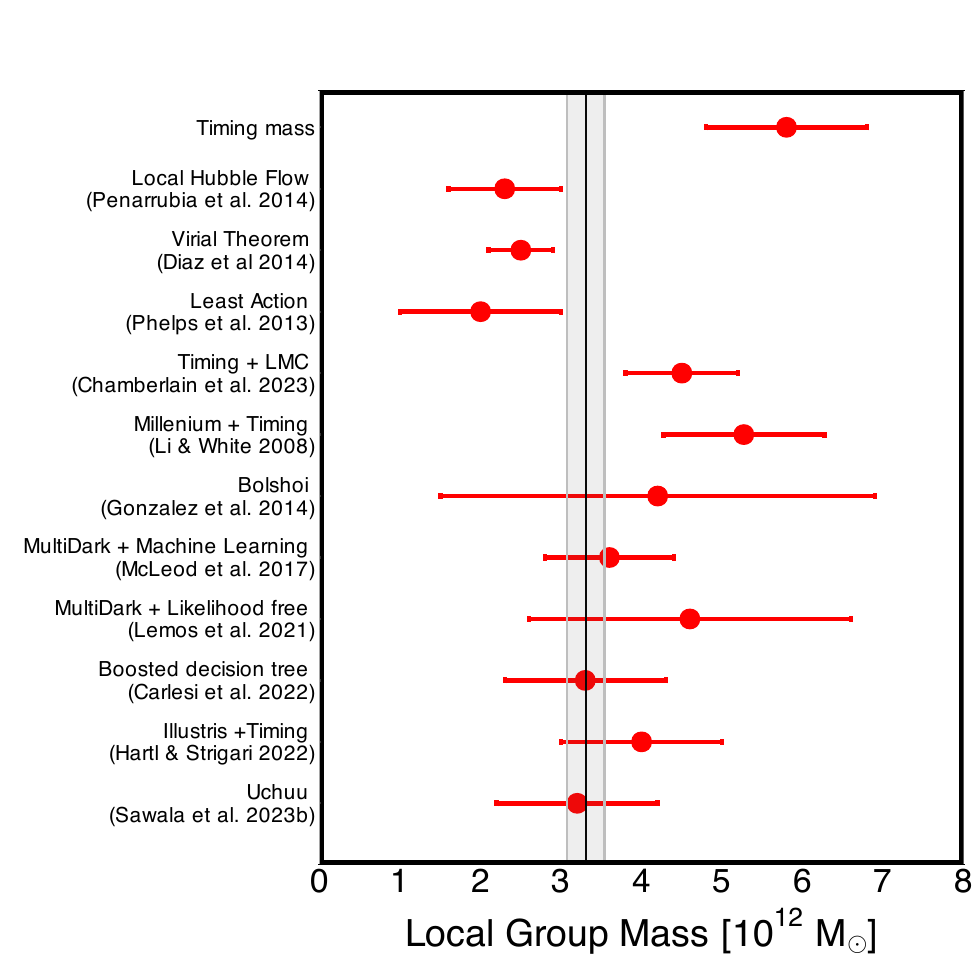} 
    \caption{Summary of mass measurements for the Local Group from different methods. The grey band shows the weighted mean and uncertainties of the measurements, $3.3 \pm 0.2 \times 10^{12}$ M$_\odot$.  
    \label{fig:LGMasses}}
\end{figure*}

 Further insight may be gained by checking the timing mass against independent LG mass estimators.~\citet{2013ApJ...775..102P} use a numerical least action method to obtain a LG mass with large uncertainties, though it is consistent with the timing mass that was derived above.~\cite{2014MNRAS.443.1688D} apply the virial theorem in combination with momentum conservation and find that the deduced mass is lower than that obtained from the timing mass. This analysis is also is subject to assumptions on the kinematics of the tracers of OLGMs used in the virial analysis. The results for these LG mass estimates are summarized in Figure~\ref{fig:LGMasses}. Refinements to the timing analysis and additional LG mass estimates are discussed in the following sections. 

\section{Local Group properties from simulations} 
\par Simulations provide the best theoretical tool for studying the properties of LG-like systems. They establish predictions for the baryonic and DM distributions within and around the MW and M31, as well as for the population of dwarf galaxies, satellite galaxies, and DM substructure within the local environment. Local Group-like systems are identified within cosmological simulations starting from a volume much larger than that which characterizes the LG itself. Identifying such LG analogues, as well as their ubiquity in the universe, starts from a selection criteria for LG-like systems. 
 
\par ~\cite{1991ApJ...376....1K} were among the first authors to study LG-like systems in simulations, and analyze them in the context of the timing mass. The full simulation of these authors was performed with a side of length 40 Mpc and contained $524,288$ particles. These were N-body simulations, which followed only the evolution of the DM under gravity within an expanding universe. Using criteria similar to that detailed below, they identify a sample of 46 LG-like systems, and showed that the masses deduced from the simple timing mass analysis was in reasonable agreement with the true masses from the simulations. 

\par The modern suites of cosmological simulations of galaxy formation are now performed within much larger volumes and follow the evolution of many more particles~\citep{Vogelsberger:2019ynw}. For the purposes of the discussion below the simulations can be classified into three general categories: 1) N-body simulations that only follow the gravitational evolution of DM particles, 2) N-body simulations that invoke semi-analytical models of galaxy formation, relying on prescriptions for the baryon content of DM halos, and 3) simulations with the full physics of galaxy formation, including baryons and DM. 

\par This section reviews the results from simulations of LG-like systems, how they are used to inform the timing mass, and more generally the properties of the LG. The discussion begins with methods for identifying LG-like systems in large-volume cosmological simulations, then moves on to discuss how LG-like systems are studied in zoom-in simulations. The section ends with a discussion of how the LG mass may be estimated with the aid of simulations. 

\subsection{Local Group analogues}
\par Local Group analogues have been identified in suites of simulations characterized by different input physics, volume, and numerical resolution. Though the detailed criteria for selecting samples of LG analogues differ, there are several general characteristics that all of the samples share. These characteristics include the DM halo masses of the MW and M31 analogue systems, the separation between the center of these two halos, and their relative velocities. It is additionally typical to exclude systems in which there is a third massive halo within a distance of a few Mpc from the barycenter of the MW-M31 analogue system. 

\par As an example application of the above criteria,~\cite{2008MNRAS.384.1459L} used the DM-only Millennium simulation to identify LG-like systems using a range of maximum circular velocities ($V_{max}$), taken as a proxy for halo mass, of $150 \, {\rm km \, s}^{-1} < V_{max} < \, 300 \, {\rm km \, s}^{-1}$. They identify a sample of $16,479$ or $23,429$ halos fitting this criteria, depending on whether or not they include an isolation criteria.~\cite{2014ApJ...793...91G} used the DM-only Bolshoi simulation, which has a side length of 357.1 Mpc, and a particle mass of $1.35 \times 10^8$ M$_\odot$. Imposing a DM halo mass cut of $5 \times 10^9 < M_\odot < 5 \times 10^{13}$ M$_\odot$, they find a sample of 4177 LG analogues at $z=0$. 
  
\par As discussed above the halo virial masses, or similarly maximum circular velocities, are difficult to directly measure, other physical properties may be used to identify LG-analogue systems. Such properties include stellar disk mass, galaxy morphology, bulge-disc morphology, satellite galaxy stellar mass function, and satellite radial distribution. Large volume simulations implementing full galaxy formation physics may be utilized to select LG analogues in terms of these more observationally-accessible quantities. Using the highest resolution, and smallest volume, Illustris TNG100 simulation~\citep{Nelson:2018uso,2024MNRAS.535.1721P}, which has a side length of 110.7 Mpc,~\cite{2020ApJ...890...27Z} identify a single LG analogue using a strict stellar mass cut criteria. Using a wider magnitude range for the MW and M31,~\cite{2022MNRAS.511.6193H} use the TNG300-1 simulation, which has a side length of 292.9 Mpc and a particle mass resolution approximately ten times greater than that of TNG100. In TNG300-1, a sample of 597 LGs were identified via similar MW and M31 stellar mass cuts, which is consistent with the results from TNG100 when scaling up the volume. 

\par Table~\ref{tab:LGanaglogues} summarizes the LG analogues identified in several large volume simulations. Shown are the samples obtained from DM-only and full physics simulations, including the samples sizes, and specific selection criteria. Depending on the specific criteria and simulation used, sample sizes range anywhere from several to upwards of $10^6$ LG analogues. 

\par Complementing the large-volume simulations described above, zoom-in simulations of the LG provide more detailed properties of the substructure in and around the MW and MW halos. Examples of two such simulation suites are APOSTLE~\citep{2016MNRAS.457..844F} and ELVIS~\citep{2014MNRAS.438.2578G}. The APOSTLE simulation suite contains 12 halo pairs re-simulated from the large volume with mass resolution over 100 times greater than that of Millennium. Formation times of LG analogues have been studied in a sample in the FIRE simulations~\citep{2020MNRAS.497..747S}. 

\par The simulations described above select analogues derived from mostly internal LG properties. Recent simulations have expanded upon this traditional selection criteria. In these simulations, LG analogues have been identified in regions of the universe constrained to produce large-scale structure around the LG~\citep{Sawala:2021npe,2024A&A...691A.348W}. ~\citet{2013ApJ...767L...5F,2016MNRAS.458..900C} select LG candidates in constrained simulations based on energy and angular momentum. As the census of the galaxy and matter distribution surrounding the LG continues to improve, these simulations and selection criteria will become more valuable for theoretical predictions of the LG properties. 

\subsection{Calibrating the timing mass} 
\par The LG samples identified in simulations may be used to calibrate the timing mass estimate. From their sample in the DM-only Millennium simulation,~\cite{2008MNRAS.384.1459L} found the timing-mass estimator to be unbiased, in the sense that it on average accurately estimates the true mass of their sample of LG analogues. These authors provide an estimate of the intrinsic, or ``cosmic", scatter in deriving the LG mass from the timing analysis. This analysis neglects the cosmological constant in the potential, and was performed before the measurement of the LG transverse velocity. Using the sample described above~\citet{2014ApJ...793...91G} extend and include the impact of the MW-M31 tangential velocity and environmental constraints and show that the timing mass overestimates the true mass by a factor $\sim 1.3-1.6$.

\par Extending the above analysis to the Illustris simulation,~\cite{2022MNRAS.511.6193H} show that the timing mass is unbiased when neglecting the cosmological constant term, but is mildly biased when the term is included. This bias is in the sense that the timing mass overestimates the true mass of the system, which is consistent with the analysis of~\cite{2013MNRAS.436L..45P} and~\citet{2014MNRAS.443.2204P}, who find that including a cosmological constant term in the equations of motion increases the deduced LG mass by $\sim 13\%$. 

\par Does the bias and scatter in the timing mass result from a flaw in the selection of the LG samples from simulations, or from inaccurate assumptions in the timing mass estimator itself? A plausible source for both the bias and the scatter is the simplified definition of mass from the timing calculation. Like~\cite{1991ApJ...376....1K}, ~\cite{2023MNRAS.526L..77S} argue that the systematic is due to the definition of MW and M31 halo mass. Following the evolution of the mass distribution of the halos, they find that the timing mass correlates with a mass within a sphere centered on the midpoint of the MW-M31 system, with a radius given by the separation between the MW and M31. This plausibly explains the bias and/or scatter for systems that have evolved in isolation and are well-described by the simple, unperturbed keplerian orbit. However, the assumption for such a simplified orbit may not be realistic, and has not yet been specifically tested in simulations. For example, this does not account for increase in angular momentum of the system resulting from tidal interactions with the nearby matter distribution, or for galaxy mergers. 
 
\par The most massive dwarf galaxies in the LG, such as the LMC and M33, induce an additional systematic in estimating the timing mass. To make a more faithful comparison between the estimated mass and the true mass, simulations may add a criteria including an LMC-like object in the system~\citep{2020MNRAS.498.2968L}. Beyond including LMC-like objects in the selection criteria, a massive third body has two distinct kinematic effects. First, a massive enough third body in the LG will induce a shift in the position of the barycenter for the MW-M31 system.~\cite{2022MNRAS.511.6193H} examine this effect in simulations using 12 LG analogs with an LMC-like object, finding a range of timing masses that is consistent with a sample without LMC-like objects. This implies that the shift due to the position of the barycenter is similar in magnitude to the shift when accounting for the cosmological constant as described above. 

\begin{table}[h]
\tabcolsep7.5pt
\caption{Local Group analogues in simulations. The first column is the simulation, the second column is the reference, the third column is the simulation volume, and the fourth column is the number of pairs identified. All rows use the mass, velocity, and isolation as selection criteria, with the exception of the last row, which uses Galaxy absolute magnitude, velocity, and isolation. 
Abbreviations: MDPL, MultiDark Plank simulation; SMDPL, Small MultiDark Planck simulation.}  
\label{tab:LGanaglogues}
\begin{center}
\begin{tabular}{@{}c|c|c|c@{}}
\hline
\hline 
Simulation & Reference & Volume&Sample \\ 
 & & (Mpc) &size \\ 
\hline
\hline 
Milenium 1 &~\citet{2008MNRAS.384.1459L} &684.9 & $2 \times 10^5$ \\ 
Bolshoi &~\citet{2014ApJ...793...91G} &357.1 &$4 \times 10^3$ \\ 
MDPL &~\citet{2017JCAP...12..034M}  &590.2 & 30,190 \\ 
SMDPL &~\citet{2021PhRvD.103b3009L}  & $1.4 \times 10^3$ & $1.1 \times 10^6$ \\ 
Uchuu &~\citet{2023MNRAS.521.4863S} & $3 \times 10^3$ & $1.4 \times 10^6$ \\ 
Illustris &~\citet{2022MNRAS.511.6193H} &292.9 & 597 \\ 
\hline
\end{tabular}
\end{center}
\end{table}

\par The second kinematic effect of an LMC-like object is discussed above, namely the addition of a travel velocity of the center of the MW disk relative to the frame defined by the MW DM halo. The resulting reflex motion may be estimated using the measured travel velocity~\citep{2021NatAs...5..251P, 2021ApJ...919..109G}, which alters both the radial and tangential MW-M31 relative velocity. To systematically account for this effect in simulations,~\cite{2022ApJ...928L...5B} inject LMC-like objects into a sample of LG-like systems from Illustris. They wind back the orbit of the LG and include an LMC into the system, and compare this evolution to one without the LMC. They find a shift in both the relative radial and tangential velocity, with a larger (downward) shift in the radial velocity, while the tangential velocity increases by less. The resulting shift is shown in Figure~\ref{fig:vrvtmass}. Both of these shifts are consistent with that deduced in~\citet{2023ApJ...942...18C}. This velocity shift increases the bias between the timing mass and the true LG mass.  

\subsection{Estimating the Local Group mass from simulations}
\par In addition to their use as a calibration for the timing mass, the LG samples obtained from cosmological simulations may be used to estimate the LG mass. With a sample of LG-analogues selected based on physical properties given in Table~\ref{tab:LGanaglogues}, the resulting sample may be interpreted as the probability for the LG mass, given such properties. In statistical language, this probability can be interpreted as the likelihood for the LG mass, given the set of properties being selected on. The resulting distribution of the likelihood as a function of halo mass then provides an estimate of the LG mass. 

\par~\cite{2017JCAP...12..034M} use  machine learning methods to find $M_{200} = 3.6 \times 10^{12}$ M$_\odot$ when using the large tangential velocity.~\cite{2021PhRvD.103b3009L} use a likelihood inference-free method to obtain $M_{200} = 4.9 \times 10^{12}$ M$_\odot$.~\cite{2022MNRAS.513.2385C} apply boosted decision trees to a sample of 2148 LG-analogues in DM only simulations to obtain a LG mass $3.31 \times 10^{12}$ M$_\odot$.~\cite{2023MNRAS.521.4863S} apply a Gaussian process regression to LG analogues in the UCHUU simulation to obtain an LG mass of $3.31 \times 10^{12}$ M$_\odot$.~\citet{2020ApJ...890...27Z} use simulations with semi-analytic models to deduce a LG mass at the higher end of this range.~\cite{2022ApJ...939...16F} introduce a selection criteria based on the barycenteric speed of the LG system, demanding that this speed be consistent with the measured LG speed with respect to the CMB. The above LG-mass estimates are statistical in nature, and depend on the selection criteria used for the LG analogues.~\cite{vanderMarel:2007yw} discuss how the assumed prior on the MW-M31 tangential velocity affects the deduced value for the LG mass. Updated measurements for the LG mass from simulations are summarized in Figure~\ref{fig:LGMasses}. 

\par The halo masses of the MW and M31 may be obtained from simulations using similar methods to obtain the LG mass.~\cite{2008MNRAS.384.1459L} apply a timing-like analysis using the MW and the satellite Leo I to estimate a MW mass of $2 \times 10^{12}$ M$_\odot$.~\cite{2011ApJ...743...40B} identify MW-like systems that have LMC-like satellite galaxies, and use this to obtain a MW virial mass of $1.2 \times 10^{12}$ M$_\odot$. For M31,~\cite{2023ApJ...948..104P} use a simulation-based method that is independent of assumptions of dynamical equilibrium of the satellite system, and obtain a M31 mass of $3 \times 10^{12}$ $M_\odot$.~\cite{2023PhRvD.107j3003V} use simulations and artificial intelligence to find results for the MW and M31 halos that are consistent with the results above.~\citet{2024OJAp....7E..50K} model the impact of the LMC on the velocity dispersion of the MW in simulations and find the MW virial mass to be $9.96 \times 10^{11}$ M$_\odot$. 

\section{Local Group environment}
\par The above discussion has focused only on the galaxies and the matter contained within the LG, and has not accounted for matter beyond the LG turnaround radius. This effectively assumes that the LG has evolved in isolation, and that the gravitational influence of matter not contained within the LG does not have an impact on it. With these assumptions a simple dynamical analysis implies conservation of energy and angular momentum, but is this assumption appropriate? Do nearby galaxies exert a tidal influence on the LG? Answering these questions involves examining the environment surrounding the LG. 

\par The LG is embedded in the so-called Lanikea supercluster, which extends across a distance of approximately 160 Mpc~\citep{2014Natur.513...71T}. Within this larger extended structure, the LG resides within a filament that stretches from the Virgo to the Fornax clusters along the supergalactic plane. On smaller scales within this filament, the MW, M31, and the galaxies within the LG are embedded within a Local Sheet, which is a flat structure extending $\sim 6$ Mpc across~\citep{2010Natur.465..565P}. The Local Sheet is dominated by  a dozen bright galaxies that are arranged around the MW and M31~\citep{2014MNRAS.440..405M}. The nearest galaxy of similar magnitude to the MW and M31 is Centaurus A, which is at a distance of $\sim 4$ Mpc from the MW. It is interesting to compare to predictions from simulations, which show that the MW local neighborhood appears to be unusual as compared to observations~\citep{2020MNRAS.494.2600N,2023MNRAS.520L..28A}.  

\par The discussion in this section focuses on the LG environment, and how it may affect the properties of the LG. The discussion begins by reviewing the Hubble flow just outside the LG, how this compares to predictions, and how it may be used to estimate the mass of the LG. The discussion then moves on to examine the possibility of preferential alignment between the LG and the surrounding environment, and the extent to which the nearby matter distribution exerts a dynamical influence on the LG. 

\subsection{Local Hubble flow}

\par The kinematics of galaxies near the boundary of the LG and just beyond the turnaround radius may be used to estimate the mass of the LG. Describing the mass distribution of the LG by a point mass centered at the LG barycenter, the orbit of a galaxy obeys the same equation that describes the timing mass, Equation~\ref{eq:equationofmotion}. Solving this equation, the velocity of a galaxy at a distance $r$ from the barycenter may be approximated by $v(r) \simeq A r - B/\sqrt{r}$~\citep{1981Obs...101..111L}, where $A, B$ are constants. This solution is also valid when including the cosmological constant, which introduces an additional term proportional to $r$ relative to the matter-only case~\citep{2014MNRAS.443.2204P}. 

\par
\begin{figure}[h]
\includegraphics[width=4.0in]{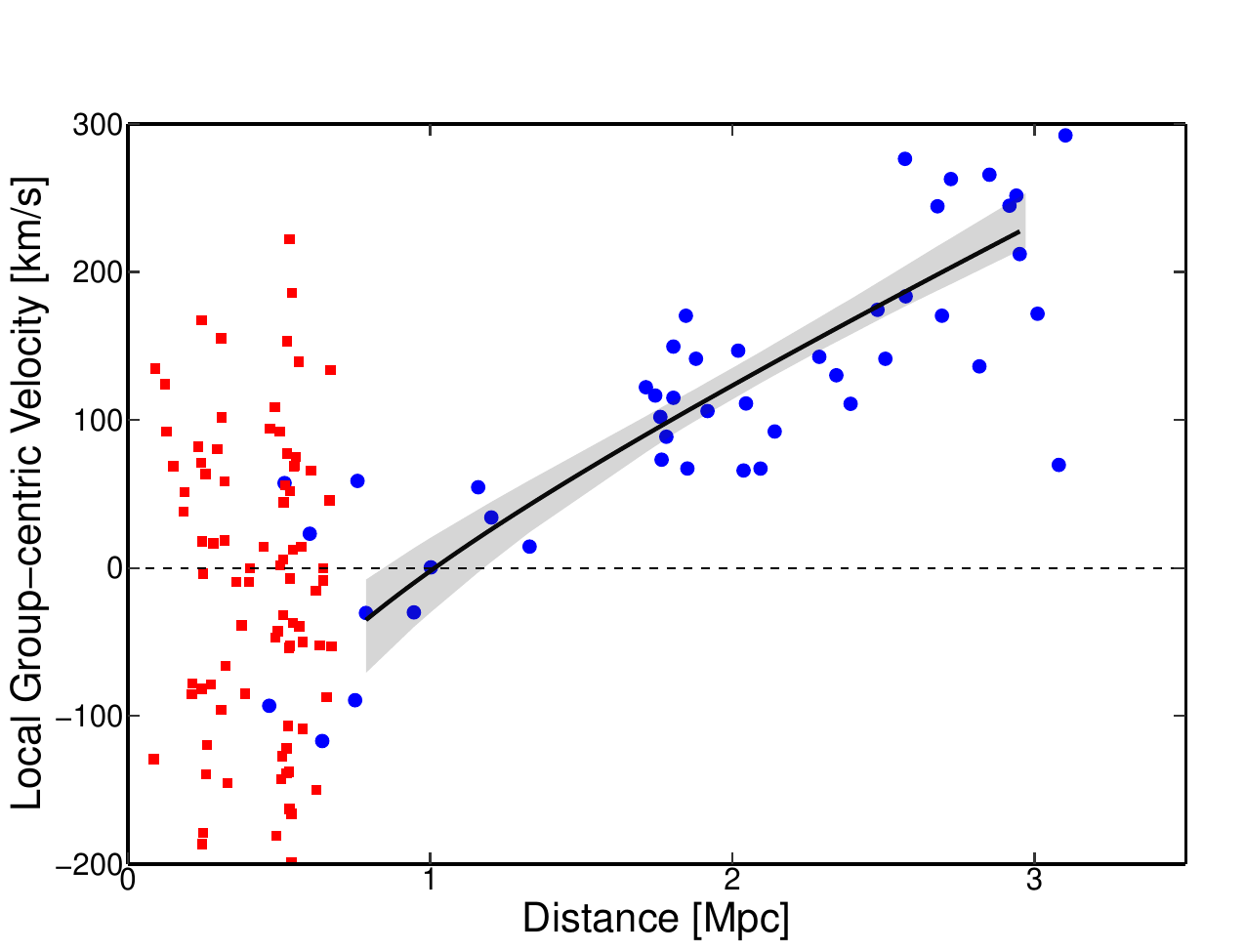}
\caption{Local Group barycentric velocity for galaxies versus distance from the Local Group barycenter. Red square points are classified as satellite galaxies of the MW or M31, while blue dots are outer LG members and galaxies beyond the LG turnaround radius. The curve shows a fit to the data in the range $0.8-3.0$ Mpc using the $v(r)$ relation given in the text, with a turnaround radius of 1 Mpc and a LG mass of $2.4 \times 10^{12}$ M$_\odot$. The grey-shaded band is derived from the confidence interval for the model parameters, and defines the uncertainty in the derived LG mass and turnaround radius. 
}
\label{fig:hubblediagram}
\end{figure}

 A simple model for galaxies near the boundary of the LG takes the entire motion to be radial relative to the LG barycenter, with no tangential component of the velocity. Under this assumption and with the above parameterization for $v(r)$, the radius at which $v = 0$ gives an estimate of the LG turnaround radius. From this turnaround radius, $R_0$, the LG mass may be estimated. Using heliocentric velocities for the population of galaxies near the LG boundary,~\cite{2009MNRAS.393.1265K} estimate a LG turnaround radius of $R_0 = 0.96 \pm 0.02$ Mpc, and a mass within the turnaround radius of $(1.88 \pm 0.18) \times 10^{12}$ M$_\odot$.~\citet{2014MNRAS.443.2204P} extend on this analysis and include uncertainties on cosmological parameters, the MW circular velocity, and the MW-M31 mass ratio to obtain a LG mass of $(2.3 \pm 0.7) \times 10^{12}$ M$_\odot$. This mass estimate is somewhat sensitive to more realistic models of the LG potential, for example including a quadrupole term to describe the MW and M31 mass distribution.  An example of this analysis is shown in Figure~\ref{fig:hubblediagram}. 

\par The dispersion about the fitted mean relation of $v(r)$ provides a measurement of the Local Hubble Flow (LHF). The measured value for this dispersion is $\sim 30$ km/s~\citep{1972ApJ...172..253S,2002A&A...389..812K},  implying a ``cold" LHF, which is in tension with predictions of this quantity from field galaxies in simulations~\citep{1997NewA....2...91G}. More direct studies of the LHF around LG analogues in simulations show that the coldness of the LHF depends on the mass of the LG, with the mean recession speeds being consistent with LG mass estimates on the lower end, $\sim 10^{12}$ M$_\odot$~\citep{2016MNRAS.457..844F}. Note also this cold LHF may be seen as indirect evidence that galaxies do not deviate too significantly from pure radial orbits. 
  
\par An important point to stress on the LHF measurement is that the galaxy census just beyond the LG is not uniform. Therefore averaging over different regions of space may lead to a simplistic and inaccurate description. Interpreting this low LHF velocity is difficult, due to incompleteness of the measurement of the local galaxy population near the turn-around radius~\citep{2024MNRAS.532.2490S}. Simulations suggest that a substantial population of LG dwarfs have not yet been detected~\citep{2020MNRAS.493.2596F}.

\subsection{Local Group angular momentum}

\par A second means by which the local environment affects the LG dynamics is by imparting angular momentum to the LG. Information is encoded within both the magnitude and direction of the angular momentum. The question of whether there is a significant amount of angular momentum in the LG has long been a subject of interest~\citep{1977MNRAS.181...37L,1978ApJ...223..426G,1982MNRAS.199...67E}. The angular momentum is contained in two components, the orbital motion and the spins of the MW and M31 DM halos. It is likely that these arise from interactions with the tidal field from galaxies beyond the LG.

\par The properties of galaxies and DM halos are predicted to correlate with the tidal field~\citep{2021MNRAS.502.5528L}. However, in the local Universe, there is not clear evidence for such a correlation.~\citet{2004ApJ...613L..41N} find that the spins of spiral galaxies preferentially align with the Supergalactic plane, which approximately defines the plane of galaxies associated with the local supercluster. This is consistent with Tidal Torque Theory (TTT)~\citep{1984ApJ...286...38W,2002MNRAS.332..325P,2009IJMPD..18..173S}, which predicts alignment of DM halos with the intermediate axis of the cosmic web~\citep{2002MNRAS.332..325P}.~\cite{2014MNRAS.440..405M} finds the spin distribution of the brightest galaxies in the Local Volume differs from the remaining galaxies in the Local Sheet. A separate analysis shows that the spins of the most massive galaxies that dominate the local angular momentum do not align in a preferential direction with respect to the Local Sheet~\citep{2023MNRAS.522.4740K}. 

\par There have been fewer studies on the alignment of the LG with the local tidal field. The LG orbital angular momentum may be derived from measurements of the relative tangential velocity. Taking the EDR3 results for the tangential velocity, the derived magnitude of the angular momentum vector is $| \log_{10} [ L_{orb} ({\rm Mpc \, km/s}) ]| = 1.88 \pm 0.1$. This is consistent with results from simulations~\citep{2015ApJ...799...45F}. The direction of the angular momentum vector makes an angle $46^\circ$ with respect to the negative SGZ axis, so it is oriented nearly between the (SGX,SGY) plane and the (negative) normal to the plane. Including the travel velocity which shifts the radial and tangential velocity as described above, the angular momentum vector then makes an angle of 27 degrees with the negative SGZ axis. Accounting for uncertainties on the velocity components, there is still a large range of allowed values for the direction of the angular momentum vector, though it is confined to a band that is normal to the MW-M31 orbital plane. 

\par It is interesting to note that for other measurements of the proper motion, the direction of the vector may differ significantly. For example, deriving the relative velocity from the OLGMs, the angular momentum vector lies at an angle approximately 8 degrees above the (SGX,SGY) plane. Including the LMC travel velocity shifts the direction of the vector to an angle 28 degrees above the (SGX,SGY) plane. Interestingly, in this case the vector points closer to the supergalactic plane, and in nearly opposite supergalactic longitude. For the case in which the velocities are derived from the M31 satellites, the corresponding angle is approximately 30 degrees above the SGZ plane. Again for these latter two methods, there is a wide uncertainty range given the uncertainties in the velocities. 

\par
\begin{figure*}[!h]
\centering 
    \includegraphics[width=5.0in]{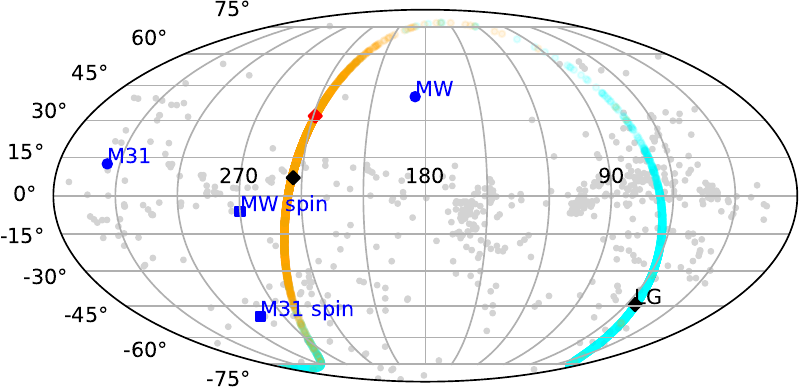} 
    \caption{Mollweide projection of the galaxy distribution in supergalactic coordinates. Light grey points are from the sample of galaxies from~\citet{2023ApJ...944...94T} with distances less than 10 Mpc. The locations of the MW and M31 are indicated, along with the direction of the spin vectors of the MW and M31. The black diamond in the lower right is the direction of the Local Group angular momentum vector determined directly from the M31 velocity, while the black diamond towards the middle left is the vector determined from the MW motion relative to the LG. The cyan points show 1000 samplings from the angular momentum distribution as determined directly from the M31 velocity, while the orange points show 1000 samplings from the angular momentum distribution as determined from the MW velocity relative to the LG. The red diamond shows the median value using the proper motion of M31 deduced from its satellite galaxies.  
    \label{fig:sg}}
\end{figure*}

 The direction of the orbital angular momentum vectors are shown in Figure~\ref{fig:sg}. Also shown in this projection is the galaxy distribution from~\citet{2023ApJ...944...94T}, including galaxies with distances $< 10$ Mpc from the Sun. In this projection, the Local Sheet is identifiable as a planar feature along nearly zero supergalactic latitude. As is seen, the angular momentum is confined to a band in these coordinates, with a direction that depends on the method used to calculate the M31-MW relative velocity. These results show that, given the ambiguity in determining the components of the MW-M31 relative velocity, there is a large uncertainty in the LG orbital angular momentum. 

\par The matter distribution within the local volume may be characterized by the principle axes of the velocity shear tensor~\citep{2012MNRAS.425.2049H}.~\citet{2015MNRAS.452.1052L} have determined the principal axes of the shear tensor in supergalactic coordinates to be in the directions $\hat e_1 =(0.3316,0.3183,-0.8881)$,  $\hat e_2 =(0.7887, 0.4229,0.4462)$, and $\hat e_3 =(-0.5175 0.8484, 0.1108)$, where the eigenvectors correspond to, respectively, the largest, intermediate, and shortest principle axis. The vector $\hat e_1$ lies nearly along the SGZ axis, which implies that the direction of maximal compression is nearly along the SGZ direction. The directions of maximal expansion in the Local Sheet correspond to nearly the (SGX,SGY) plane. 

\par Again using the measured tangential velocities, the direction of the angular momentum vector relative to this shear field may be determined. Using the EDR3 results, the projection of the LG orbital angular momentum onto the components of the shear tensor is 
    $\hat e_1 \cdot \hat L = 0.91_{-0.23}^{+0.08}$, 
    $\hat e_2 \cdot \hat L = 0.25_{-0.35}^{+0.30}$, and 
    $\hat e_3 \cdot \hat L = 0.19_{-0.29}^{+0.25}$. 
This implies that the angular momentum vector makes an angle $\sim 65$ degrees relative to the $(\hat e_2, \hat e_3)$ plane. The angular momentum vector does not lie in the $(\hat e_2, \hat e_3)$ plane, as is predicted by TTT. This is also true for the angular momentum vector derived from the OLGMs, for which the angular momentum vector is nearly equidistant between the $\hat e_1$ axis and the $(\hat e_2, \hat e_3)$ plane. 

\par Satellite galaxies that orbit their hosts are also predicted to align with the larger scale mass distribution. Many of the satellites that orbit the MW, M31, and Centaurus A, are arranged in planes, and nearly all of these planes are aligned with the principal axis describing the local matter distribution~\citep{2015MNRAS.452.1052L}. This is consistent with the results in high resolution LG simulations, which show that satellites accrete along the direction of slowest collapse~\citep{2013ApJ...767L...5F,2016MNRAS.458..900C}.~\cite{Dupuy:2022jjj} shows that satellites accrete along the direction of slowest collapse for a sample of 3 high resolution LGs, indicating that mergers may also affect alignments in the Local Volume. More generally,~\cite{2020MNRAS.491.1531C} consider the mass assembly history for the LG. They find that the median constrained merging histories for M31 and MW lie above the standard ones at the $1-\sigma$ level, similar to the earlier studies of~\cite{2011MNRAS.417.1434F}. 

\section{Summary and Discussion}

\subsection{Timing mass, Local Group, MW, and M31 masses}

\par Two questions were posed in the introduction of this article: 1) Is the sum of the  MW and M31 masses consistent with the mass measured for the LG as a whole? 2) What is the dynamics of the MW-M31 system? The first question was addressed by appealing to unique LG mass estimators, with a particular focus on the timing mass, and comparing to the MW and M31 mass estimates obtained from tracer populations extending to large radii. From the latter mass measurements, the sum of the individual masses of the MW and M31 is $M_{MW} + M_{M31} = (2.5 \pm 0.8) \times 10^{12}$ M$_\odot$. However, the weighted average of all the LG mass estimates shown in Figure~\ref{fig:LGMasses} is $M_{LG} = (3.3 \pm 0.2) \times 10^{12}$ M$_\odot$. 

\par It is interesting to examine the LG mass estimates in more detail, in particular comparing those that use input from simulations to those based purely on analysis of the observed data sets. For the simulations-based measurements, the weighted average LG mass is $(3.9 \pm 0.4) \times 10^{12}$ M$_\odot$, while for those based purely on observations, excluding the timing mass, the weighted average LG mass is $(2.8 \pm 0.3) \times 10^{12}$ M$_\odot$ [when the timing mass is included, the corresponding LG mass is $(3.0 \pm 0.3) \times 10^{12}$ M$_\odot$)]. Therefore the simulations-based measurements introduce a prior that weights towards larger LG masses, increasing the difference between the measured LG mass and the sum $M_{MW} + M_{M31}$, and introducing some tension between the two types of measurements. 

\par There is not an obvious argument for which LG mass estimate to prefer. On the one hand, for the simulations-based measurements, the estimate depends on the assumed prior for the selection of LG-like systems. On the other hand for the observations-based measurements, biases in the mass reconstruction must be properly accounted for. For example, in the case of the LHF measurement a nearly radial velocity distribution is assumed, and in the case of the virial theorem analysis, isotropy of the tracer population is assumed. Additionally for the observation-based estimates, there is still the question of the origin of the significant scatter between the different LG mass estimates. In particular, the LHF and virial mass measurements are systematically lower than the timing mass estimate. As has been shown, uncertainty in the parameters that determine the timing mass, such as the MW-M31 relative velocity and the MW circular velocity, may reduce the mass estimate relative to canonical assumptions for these model parameters. However, appealing to these changes alone it is still difficult to completely bring the mass estimates in line with the LHF and virial estimates. 

\par Considering possibilities of systematic bias in the timing mass, it is reasonable to ask whether the orbit differs from the assumed simple, isolated 2-body model. This difference may be attributed to ``internal" effects in which a massive body within the LG is altering the dynamics of the system, or ``external" effects in which the local environment is perturbing the orbit. Internal effects due to the LMC have been extensively explored, and as discussed result in a shift in the relative MW-M31 velocity. More general effects from the DM substructure in the LG may be accounted for by calibrating to simulations, which provide inconclusive results. Indeed, the bias between the timing mass and the true LG mass in simulations is negligible for an assumed matter-dominated universe, but is more significant for dark energy dominated models. The interplay between the effect of substructure and the large-scale cosmological model may therefore be more complicated than naively appears. 

\par External effects may be important either in the form of mass exerting a significant gravitational influence on the LG, through the imparting of angular momentum on the LG, or through some combination of these. The population of luminous galaxies within the Local Volume is well-characterized, however, for determining the influence on the LG the most relevant property of these galaxies is the dark-to-luminous mass ratio. It is possible that improving the population of faint and low-mass galaxies may provide more information on the matter distribution in the Local Volume, as the census of dwarf galaxies in and around the LG is still incomplete. 

\par So what conclusions can be drawn at this point? Is the timing mass biased? Though this is perhaps the easiest answer to a simple yet subtle question, it would be remiss to ignore the successes of the model. For example, the model accurately predicts that the MW and M31 are the dominant components of mass in the LG. Further, why does the estimator work so successfully for matter-dominated universes? And as a follow up why does the introduction of the cosmological constant, which must strictly be included in the orbit analysis, seemingly introduce a larger bias when calibrating to simulations? Questions such as these must be fully addressed before abandoning the timing mass as a tool for studying the LG. 

\subsection{Future measurements}
\par \par What can be learned in the future on the mass and the orbit of the LG? The kinematics of LG galaxies will certainly be improved upon, with an important example being the M31 proper motion. Aside from improving upon the kinematics of known galaxies, new measurements will be important, as an example the proper motions of the galaxy population within the LG. These represent the most basic of the many possible kinematic measurements that can be made to improve understanding of the dynamics of the LG. 

\subsubsection{M31 tangential motion}

\par Direct measurements of the M31 proper motion are expected to improve with future astrometric data sets. Proper motion measurements with {\it Gaia} DR2 and EDR3 relied on the sample of red and blue supergiants, a bright population that is localized in the M31 disk. The upcoming {\it Gaia} DR4 will provide improved astrometry, photometry and spectroscopy for all stars, as well as epoch-level astrometry. This will expand the sample of member stars and provide a better characteristic of the binary star population. For example, even a reduction in the error on the proper motions by a factor of two would imply an improved measurement of the tangential velocity, and thereby an improved understanding of the MW-M31 orbit. 

\subsubsection{Massive satellites}

\par The impact of the LMC on the dynamics of the MW and the LG has been studied at great length in the recent literature. It is possible the M33 had a similar effect on M31, which has yet to be studied in detail. This is in part because of the lack of reliable measurement for the M33 proper motion. This is also likely to improve with upcoming Gaia DR4 and DR5 data releases, which will allow future analyses to account for the ``travel" velocity induced by M33 on M31. 

\subsubsection{Satellites galaxies and Outer Local Group members}

\par As emphasized it is important to establish a consistency between the direct astrometric and the indirect measurements of the M31 tangential velocity. This may be accomplished with definitive measurements of the proper motions for OLGMs~\citep{2021MNRAS.501.2363M}. Though not sufficiently within reach for current {\it Gaia} and HST measurements, it presents an important measurement opportunity for {\it Gaia} DR4 and DR5. This is similar to the recent improvement in proper motions for satellite galaxies through HST and {\it Gaia}, which are precise enough to estimate not only the orbits, but also possibly the mass distributions from internal proper motions~\citep{Strigari:2007vn,Strigari:2018bcn,Guerra:2021ppq} and the structural~\citep{2020MNRAS.495.4124F,2021ApJ...908..244D,2022MNRAS.512.5601Q,2023MNRAS.523..123S} and orbital~\citep{2022ApJ...940..136P} parameters of dwarf galaxies. Improved kinematics of LG galaxies may also shed light on interesting possible features in the distribution of galaxies in the  LG~\citep{2005ApJ...629..219Z,2013MNRAS.435.1928P,2018MNRAS.473.4033B}.  

\subsection{Dark matter, dark energy, and the local expansion rate} 
\par In the analysis of the orbit equation above, the sensitivity of the LG dynamics to the assumed cosmological model has been emphasized, in particular the local expansion rate, and the dark matter and dark energy densities. The prospects for using the LG dynamics to constrain dark energy has been considered~\citep{2014MNRAS.443.2204P,2017MNRAS.466.4813C,2020JCAP...09..056M,Benisty:2023vbz,2024A&A...689L...1B}, though the LG dynamics determine cosmological parameters with far less precision than more classical cosmological probes. Though interestingly, the LG does allow for the possibility of probing very local values of cosmological parameters, and possibly even anisotropy in measured values. A precise measurement of the LG orbit may also be sensitive to post-newtonian orbital corrections and modified gravity~\citep{Benisty:2023ofi,Benisty:2023clf}. These topics highlight the unique status of the LG as the boundary at which the effect of cosmology at large-scale begins to become manifest. 

\section*{ACKNOWLEDGMENTS}
The author acknowledges support from the DOE Grant No. DE-SC0010813 and from the Texas A\&M University System National Laboratories Office and Los Alamos National Laboratory. We thank Joss Bland-Hawthorn, Marla Geha, and Andrew Pace for comments and feedback on an earlier draft the article, and Odelia Hartl for reviewing this article and collaboration on this topic. 

\bibliography{bibliography}%

\end{document}